\documentclass[lettersize,journal]{IEEEtran}
\usepackage{amsmath, amssymb, amsfonts}
\usepackage{amsthm}
\usepackage{bm} 

\usepackage[ruled,vlined]{algorithm2e}
\usepackage{algorithmic}

\usepackage{array}
\usepackage[caption=false,font=normalsize,labelfont=sf,textfont=sf]{subfig}
\usepackage{textcomp}
\usepackage{booktabs}
\usepackage{stfloats}
\usepackage{url}
\usepackage{verbatim}
\usepackage{graphicx}
\usepackage{cite}
\usepackage{subfig}
\usepackage{float} 
\usepackage{amssymb}
\usepackage{tikz}      
\usepackage{xcolor}    
\usepackage{graphicx}  
\usepackage[colorlinks=true,citecolor=purple,linkcolor=purple,urlcolor=purple]{hyperref}

\newcommand{\rev}[1]{#1}
\newcommand{\polish}[1]{#1}

\begin{document}

\title{A High-Resolution TOA Estimation Algorithm Based on Compressed Sensing for Coherent Multi-Band Signals}

\author{Jialun Kou,~\IEEEmembership{Student Member,~IEEE,}
        Marc Bauduin,~\IEEEmembership{Member,~IEEE,}
        Achiel Colpaert,~\IEEEmembership{Member,~IEEE,}
        Zhuangzhuang Cui,~\IEEEmembership{Member,~IEEE,}
        Sofie Pollin,~\IEEEmembership{Senior Member,~IEEE,}

\thanks{
This work is supported by the SUNRISE-6G and MultiX projects, funded by the Smart Networks and Services Joint Undertaking (SNS JU) under the European Union’s Horizon Europe research and innovation programme, with Grant Agreement Nos. 101139257 and 101192521. (Corresponding author: Sofie Pollin)}
\thanks{ Jialun Kou, Achiel Colpaert, Zhuangzhuang Cui, and Sofie Pollin are with WaveCoRe, Electrical Engineering Department (ESAT), KU Leuven,
3001 Leuven, Belgium. (e-mail: {firstname.lastname}@kuleuven.be).} 
\thanks{ The authors are also with imec, Kapeldreef 75, 3001 Leuven, Belgium.}
}



\maketitle

\begin{abstract}
High-resolution time-of-arrival (ToA) estimation is crucial for accurate ranging and localization in future wireless networks. While multi-band operation can synthesize a large virtual bandwidth by probing several separated subbands, \polish{the resulting non-contiguous
frequency layout} induces pronounced sidelobes and dense local maxima in the delay response, making reliable ToA estimation challenging.
This paper studies coherent multi-band ToA estimation from a single multi-band \polish{channel state information (CSI)} snapshot.
\polish{We derive a multi-band least-squares formulation and analyze its point-spread function (PSF) and Cram\'er--Rao bound (CRB) to characterize the effects of the
subband layout on delay accuracy and sidelobe ambiguity.}
Building on this analysis, we propose a low-complexity two-stage particle swarm optimization estimator and incorporate it into orthogonal matching pursuit with a RELAX-type cyclic refinement procedure for multipath delay estimation, mitigating basis mismatch and improving robustness to the dense local maxima induced by \polish{non-contiguous multi-band layouts}.
\polish{Numerical results are consistent with the PSF and CRB analyses and show that
the proposed framework approaches the CRB in the single-path setting and
achieves accurate multipath delay estimation with low computational
complexity. Real channel measurements further validate the
predicted impact of different subband layouts on delay-estimation performance.}
\end{abstract}


\section{Introduction}

\begin{figure*}[t]
\centering
\includegraphics[width=0.95\textwidth]{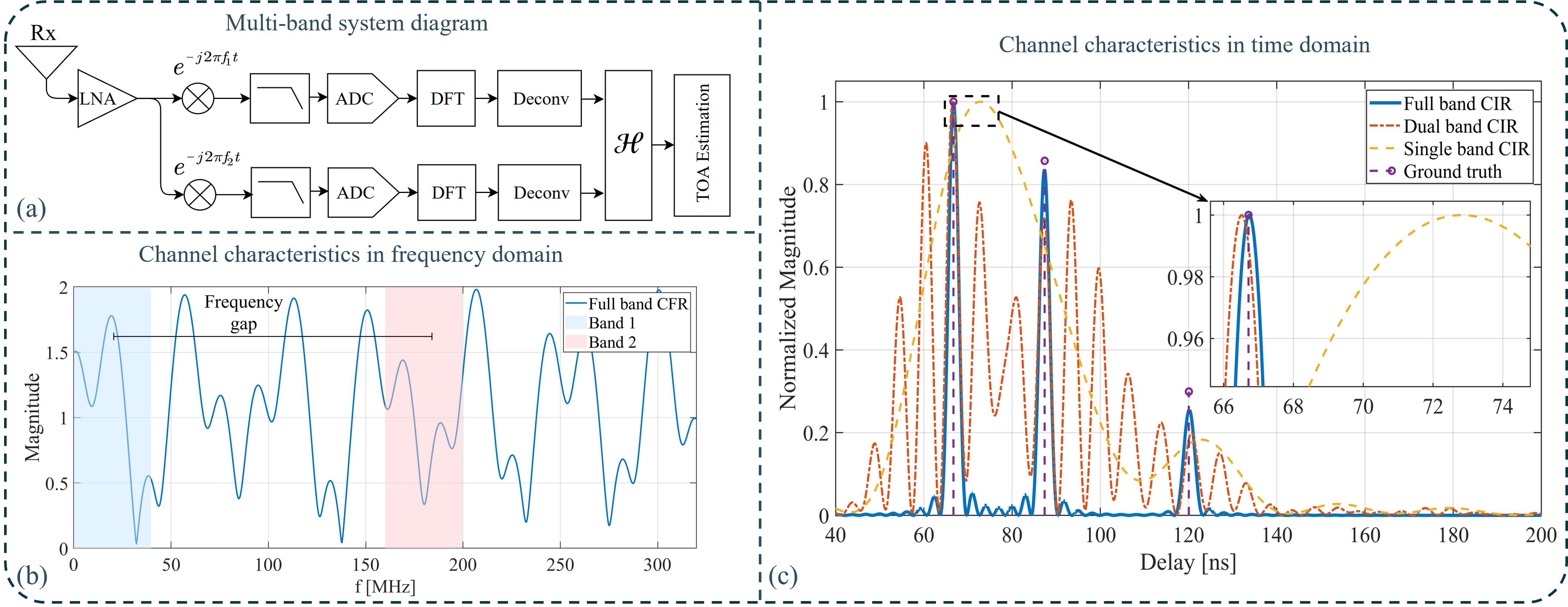}
\caption{Multi-band system diagram and channel characteristics. (a) System diagram. (b) Frequency response for two 40 MHz subbands selected from a 320 MHz wideband, separated by a 160 MHz gap. (c) Channel impulse responses obtained from the full band, the non-contiguous multi-band, and a single subband, together with the ground-truth path delays.}
\label{fig:dualband_illustration}
\end{figure*}
\IEEEPARstart{F}{uture} wireless networks are expected to deliver not only high-throughput communication but also enhanced sensing and localization capabilities. This vision has motivated integrated sensing and communication (ISAC), where sensing functions are realized by reusing existing communication infrastructure and waveforms \cite{Liufan,Adham,Xiong2026Fresnel,Haoqiu}. In line with this trend, the ITU-R IMT-2030 (6G) framework includes ISAC as one of the key usage scenarios. ISAC-enabled capabilities are anticipated to support a broad range of emerging services, such as healthcare innovations \cite{Haoqiu}, smart cities \cite{smartcity}, and autonomous driving \cite{Yunxin}.

Wireless localization is an important capability in ISAC systems that enables many location-dependent applications.
Accurate localization typically relies on estimating propagation parameters for each propagation path component, including the complex amplitude, time-of-arrival (ToA), Doppler shift, and angle-of-arrival (AoA). Among these, ToA-based ranging has been widely studied \cite{YassinSurvey2017}, where the range is obtained by multiplying the ToA of the line-of-sight (LoS) (or first-arriving) path by the speed of light.
In conventional receivers, ToA is estimated by correlating the received signal with a known reference signal \cite{intro_delayest}, and the achievable delay resolution is primarily determined by the effective signal bandwidth. In multipath environments, limited bandwidth leads to insufficient delay resolution, so that the LoS path cannot be reliably separated from closely spaced multipath components (MPCs), resulting in biased range estimates and degraded localization performance \cite{UWB_ACM}. Increasing the signal bandwidth improves delay resolution and has motivated the use of ultra-wideband (UWB) technology for localization \cite{UWB_ACM}. However, UWB systems are typically restricted to relatively small coverage areas due to stringent transmit-power limits, and they require high-rate analog-to-digital converters (ADCs) to capture wideband signals. To mitigate these limitations, modern wireless systems increasingly exploit non-contiguous spectrum to synthesize a larger virtual bandwidth and hence enhance delay resolution. For instance, IEEE~802.11be (Wi-Fi~7) supports flexible channelization with bandwidths of 20, 40, 80, and 160~MHz and enables the activation of multiple separated channels via multi-link operation (MLO) \cite{WIFI7}. This allows multi-band observations to be jointly processed for localization while keeping the instantaneous bandwidth and the required sampling rate moderate on each channel.

A typical multi-band ToA estimation scheme is shown in
Fig.~\ref{fig:dualband_illustration}a. We focus on coherent multi-band
processing, which jointly exploits the common path-delay information and the relative phase relationships across the selected subbands
~\cite{subspace3}.  Under this model, the main difficulty of ToA estimation stems from the non-contiguous frequency layout. Combining multiple separated subbands increases the effective frequency aperture and can sharpen the mainlobe of the delay response, whereas the inter-band gaps induce pronounced sidelobes and multiple local maxima, which may lead to spurious paths and unreliable delay estimates. Figs.~\ref{fig:dualband_illustration}b and \ref{fig:dualband_illustration}c illustrates this effect for a configuration with two separated subbands. The full-band channel impulse response (CIR) yields a dominant peak close to the ground-truth LoS delay. With only a single narrow band, the limited bandwidth produces a broad mainlobe whose peak is more easily biased by MPCs and noise. In contrast, the non-contiguous multi-band CIR exhibits a much sharper mainlobe due to the enlarged effective aperture, but the frequency gap generates strong sidelobes and multiple false peaks around the true delay. This sidelobe-induced delay ambiguity is the central challenge addressed in this work.

Existing multi-band ToA estimation techniques can be broadly grouped into maximum-likelihood (ML), subspace-based, and compressed-sensing (CS) approaches~\cite{SAGE1,SAGE2,twoStage,subspace1,subspace2,subspace3,Decimeter,dunSparseSignalBands2020,wifiomp,hisac}. ML and ML-inspired schemes include a low-complexity model-based estimator that reconstructs an equally spaced tap channel and detects the first path by minimizing energy leakage~\cite{SAGE1}. In addition, the space-alternating generalized expectation-maximization (SAGE) framework has been extended to fuse multi-band ToA information, which improves ranging accuracy but relies on iterative procedures that are sensitive to initialization and may converge to local optima~\cite{SAGE2}. A two-stage global estimation scheme has also been proposed to more effectively exploit multi-band gains, combining a coarse Bayesian inference stage with a refined global search. While this approach achieves high accuracy, its message passing and population-based optimization over a high-dimensional parameter space lead to high computational complexity~\cite{twoStage}.

Subspace-based methods achieve super-resolution by organizing multi-band channel samples into structured data matrices and exploiting shift-invariance relations to estimate path delays via ESPRIT-type or subspace-fitting algorithms~\cite{subspace1,subspace2,subspace3}. 
These methods rely on accurate estimation of the signal and noise subspaces. In practice, acquiring multiple \polish{channel state information (CSI)} snapshots may be challenging due to overhead and channel time variations. 
To enable operation in the single-snapshot regime, representative approaches such as the Multiband Weighted Delay Estimation (MBWDE) algorithm employ data-extension strategies to construct a structured low-rank representation from a single CSI snapshot and perform subspace decomposition accordingly~\cite{subspace2}. 
Nevertheless, such single-snapshot subspace estimates are generally more sensitive to noise and model mismatch, which can perturb the estimated subspaces and degrade the ToA accuracy, especially in \polish{low signal-to-noise ratio (SNR)} or rich-multipath scenarios.

Compared with subspace-based approaches, CS-based methods explicitly exploit channel sparsity and can be naturally applied to a single CSI snapshot by solving a sparse inverse problem, which makes them appealing when only limited measurements are available. One representative approach formulates multi-band WiFi ToA estimation as a sparse non-uniform inverse Fourier reconstruction problem and uses $\ell_1$-regularized recovery of the CIR to achieve decimeter-level localization with a single access point~\cite{Decimeter}. 
A related approach adopts a similar sparse non-uniform inverse Fourier recovery, but augments it with a Cram\'er--Rao Bound (CRB) based band selection strategy in order to reconstruct a high-resolution multipath CIR while keeping the occupied spectrum limited~\cite{dunSparseSignalBands2020}. 
Other works rely on sparse reconstruction based on orthogonal matching pursuit (OMP) operating on multi-band \polish{channel frequency response (CFR)} measurements, which enables sparse high-resolution CIR estimation for WiFi localization and few-centimeter range sensing in multi-band ISAC systems~\cite{wifiomp,hisac}. 
Compared with non-convex ML-type iterative schemes, CS formulations are less prone to local-optimum convergence. Moreover, CS methods have been shown to outperform subspace-based approaches in noisy environments, particularly when strong dominant clutter or LoS components are present~\cite{advantageOFCS}.

A key challenge for CS-based ToA estimation is the basis mismatch problem, where the true delays do not align with the discretized delay grid used to construct the dictionary~\cite{offgrid1}. 
Using a finer grid reduces mismatch but substantially increases computational cost, which often hinders real-time processing~\cite{offgrid2}. Representative off-grid CS approaches include Newtonized OMP (NOMP)~\cite{NOMP} and various interpolation-based refinements~\cite{offgridOMP1,offgridOMP2,offgridOMP3,OffgridOMP4,interpolation1}, 
which mitigate basis mismatch by off-grid refinement of the delays detected by OMP. These refinements are particularly effective for uniformly spaced frequency measurements with a well-behaved Dirichlet-type response. However, in multi-band settings with inter-band gaps, the delay objective becomes strongly multimodal and may exhibit dense secondary peaks with comparable amplitudes near the true delay.
As a result, local refinement can converge to a nearby secondary peak unless the coarse detection stage provides a sufficiently accurate initialization, which typically requires a much finer coarse grid and hence higher complexity. Motivated by these limitations, we develop a low-complexity continuous-domain delay estimation framework tailored to non-uniform multi-band frequency layouts.
Specifically, we propose a two-stage particle swarm optimization (PSO) estimator that exploits the uniform spectrum of a single contiguous subband to obtain a reliable coarse delay estimate, and then refines it by maximizing the full multi-band LS objective over a confined interval, thereby substantially narrowing the PSO search region. We further incorporate two-stage PSO into sparse multipath recovery with OMP and a RELAX-type cyclic refinement procedure, resulting in the proposed PSO-OMP and PSO-RELAX algorithms for multi-band multipath delay estimation.

Simulation results show that the proposed methods achieve near-CRB accuracy and consistently outperform multi-band discrete OMP~\cite{hisac} and the subspace-based MBWDE approach~\cite{subspace2}.
Moreover, PSO-RELAX attains performance comparable to NOMP~\cite{NOMP} while requiring substantially lower execution time in our multi-band implementation.

The main contributions of this paper are summarized as follows. \begin{enumerate} \item We establish a multi-band LS formulation for delay estimation and analyze the associated \polish{point-spread function (PSF)} and CRB. The analysis quantifies how the number of subbands, the inter-band spacing, and the per-subband bandwidth jointly shape the delay-domain objective and the achievable accuracy. 
\item We propose a two-stage PSO (TSPSO) estimator for off-grid single-path delay estimation under non-contiguous multi-band frequency layouts. TSPSO combines a low-complexity coarse search on a single contiguous subband with a continuous-domain refinement of the full multi-band objective over a confined interval, achieving near-CRB performance with reduced computational cost.
\item We incorporate TSPSO into OMP and a RELAX-type cyclic refinement procedure to obtain PSO-OMP and PSO-RELAX for continuous-domain multipath delay estimation. \polish{We provide complexity analysis and validate the proposed framework through simulations and real channel measurements. The results demonstrate improved estimation accuracy and reduced computational cost compared with representative multi-band compressed-sensing and subspace-based methods.}
 \end{enumerate}

\section{Signal Model and Multi-band Objective}
\label{sec:signal_model}

\subsection{\texorpdfstring{\polish{Single-path LS model and delay-domain objective}}{Single-path LS model and delay-domain objective}}
\label{sec:single_path_ls}

\polish{
To analyze how the placement of the measured frequencies affects delay
sensitivity and sidelobe behavior, we first consider a coherent single-path
multi-band CFR model. Multipath channels are discussed in
Section~\ref{sec:coherent_multipath_estimation}, and non-coherence across
subbands is considered in Section~\ref{sec:interband_noncoherence}.

Measurements are collected from \(T\) subbands. Let \(F_t\) denote the carrier
frequency of the \(t\)th subband. Each subband provides \(M_b\) measured
frequencies with spacing \(\Delta f\). 
We define the intra-subband index set \(\mathcal{U}=
\{m-\tfrac{M_b-1}{2}\mid m=0,\ldots,M_b-1\}\), the set of \(M=TM_b\)
measured frequencies is
\(\mathcal{F}=\{f_{t,m}\mid f_{t,m}=F_t+m\Delta f,\
t=1,\ldots,T,\ m\in\mathcal{U}\}\).

Collecting the CFR samples at these measured frequencies gives the single-path observation model
\begin{align}
\mathbf{y}
&=
\alpha_0\mathbf{a}(\tau_0)+\mathbf{v},
\qquad
\mathbf{v}\sim\mathcal{CN}(\mathbf{0},\sigma^2\mathbf{I}_M).
\label{eq:vec_model_single}
\end{align}

where \(\alpha_0\) and \(\tau_0\) denote the complex path gain and path delay,
respectively, \(\sigma^2\) is the noise variance per measured CFR sample, and
the steering vector is \(\mathbf{a}(\tau)=[e^{-j2\pi f_{t,m}\tau}]_{t,m}^{\top}\).
To jointly estimate \(\alpha_0\) and \(\tau_0\), we consider the LS residual
\begin{align}
R(\alpha,\tau)
&=
\left\|\mathbf{y}-\alpha\mathbf{a}(\tau)\right\|^2.
\label{eq:joint_ls_generic}
\end{align}

For a fixed candidate delay \(\tau\), the complex gain that minimizes
\(R(\alpha,\tau)\) is obtained by the LS projection
\begin{align}
\widehat{\alpha}(\tau)
&=
\frac{\mathbf{a}(\tau)^H\mathbf{y}}
{\mathbf{a}(\tau)^H\mathbf{a}(\tau)}.
\label{eq:ls_gain_single}
\end{align}
Substituting \(\widehat{\alpha}(\tau)\) into
\eqref{eq:joint_ls_generic} gives
\begin{align}
\min_{\alpha}R(\alpha,\tau)
&=
\left\|\mathbf{y}\right\|^2
-
\frac{\left|\mathbf{a}(\tau)^H\mathbf{y}\right|^2}
{\mathbf{a}(\tau)^H\mathbf{a}(\tau)}.
\label{eq:concentrated_ls_generic}
\end{align}
Since \(\left\|\mathbf{y}\right\|^2\) is independent of \(\tau\), minimizing
the residual is equivalent to maximizing the concentrated LS objective
\begin{align}
\mathcal{C}(\tau)
&=
\frac{\left|\mathbf{a}(\tau)^H\mathbf{y}\right|^2}
{\mathbf{a}(\tau)^H\mathbf{a}(\tau)}.
\label{eq:objective_def_generic}
\end{align}

Substituting \eqref{eq:vec_model_single} into the numerator of
\eqref{eq:objective_def_generic} gives
\begin{align}
\mathbf{a}(\tau)^H\mathbf{y}
&=
\alpha_0\mathbf{a}(\tau)^H\mathbf{a}(\tau_0)+\eta,
\notag\\
\eta
&=
\mathbf{a}(\tau)^H\mathbf{v}
\sim\mathcal{CN}(0,\sigma^2M).
\label{eq:coherent_projection_single}
\end{align}
The deterministic inner product in
\eqref{eq:coherent_projection_single} is the coherent delay kernel
\begin{align}
h(\delta_\tau,\mathcal{F})
&=
\mathbf{a}(\tau)^H\mathbf{a}(\tau_0)
\notag\\
&=
\sum_{t=1}^{T}\sum_{m\in\mathcal{U}}
e^{j2\pi f_{t,m}\delta_\tau},
\qquad
\delta_\tau=\tau-\tau_0.
\label{eq:coherent_delay_kernel_definition}
\end{align}
Substituting \eqref{eq:coherent_projection_single} and
\eqref{eq:coherent_delay_kernel_definition} into
\eqref{eq:objective_def_generic} and taking expectation over the noise gives
\begin{align}
\mathbb{E}_{\mathbf{v}}\big[\mathcal{C}(\tau)\big]
&=
\frac{|\alpha_0|^2}{M}|h(\delta_\tau,\mathcal{F})|^2+\sigma^2
\notag\\
&=
\frac{|\alpha_0|^2}{M}P(\delta_\tau,\mathcal{F})+\sigma^2.
\label{eq:psf_definition}
\end{align}
Since every entry of \(\mathbf{a}(\tau)\) has unit magnitude,
\(\mathbf{a}(\tau)^H\mathbf{a}(\tau)=M\). The quantity
\(P(\delta_\tau,\mathcal{F})=|h(\delta_\tau,\mathcal{F})|^2\) is defined as
the point-spread function.

Equation~\eqref{eq:psf_definition} shows that, up to an additive noise floor,
the deterministic component of the expected objective is proportional to
\(P(\delta_\tau,\mathcal{F})\).
The PSF therefore determines the mainlobe width, which reflects local delay
sensitivity, and the sidelobe structure of the LS objective. The following
analysis studies how the subband layout shapes these PSF properties.
}

\begin{figure*}[!t]
\centering
\begin{minipage}[t]{0.32\linewidth}
    \centering
    \includegraphics[width=\linewidth]{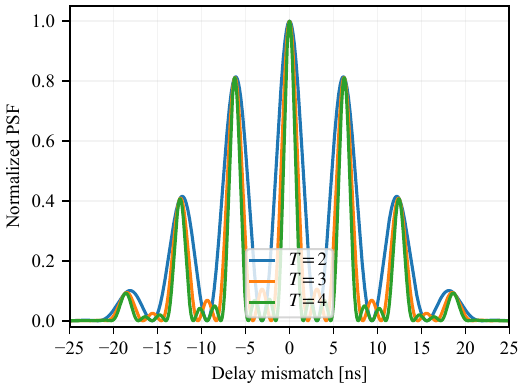}\\[-1mm]
    {\footnotesize\polish{(a) $f_{\mathrm{gap}}=160~\text{MHz}$, $M_b\Delta f=40~\text{MHz}$}}
\end{minipage}
\hfill
\begin{minipage}[t]{0.32\linewidth}
    \centering
    \includegraphics[width=\linewidth]{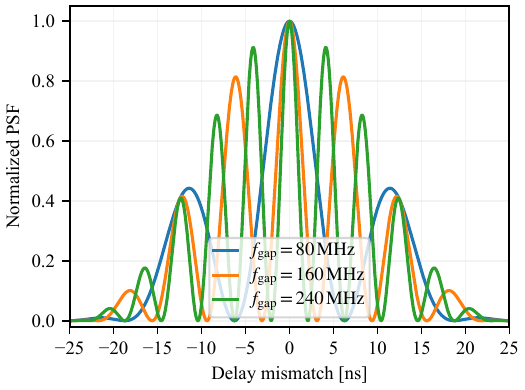}\\[-1mm]
    {\footnotesize\polish{(b) $T=2$, $M_b\Delta f=40~\text{MHz}$}}
\end{minipage}
\hfill
\begin{minipage}[t]{0.32\linewidth}
    \centering
    \includegraphics[width=\linewidth]{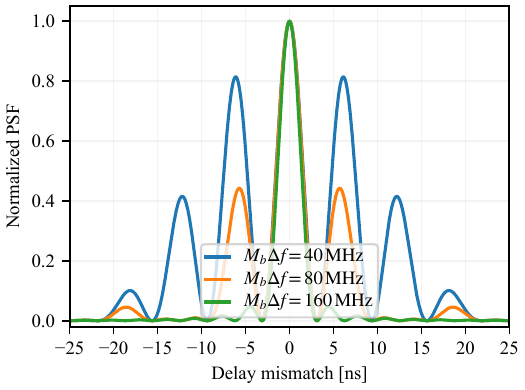}\\[-1mm]
    {\footnotesize\polish{(c) $f_{\mathrm{gap}}=160~\text{MHz}$, $T=2$}}
\end{minipage}

\caption{\polish{Multi-band PSFs under different configurations.
(a) Varying the number of active subbands $T$.
(b) Varying the inter-band frequency gap $f_{\mathrm{gap}}$.
(c) Varying the per-subband bandwidth $M_b\Delta f$.}}
\label{fig:PSF}
\end{figure*}

\begin{figure*}[!t]
\centering
\begin{minipage}[t]{0.335\linewidth}
    \centering
    \includegraphics[width=\linewidth]{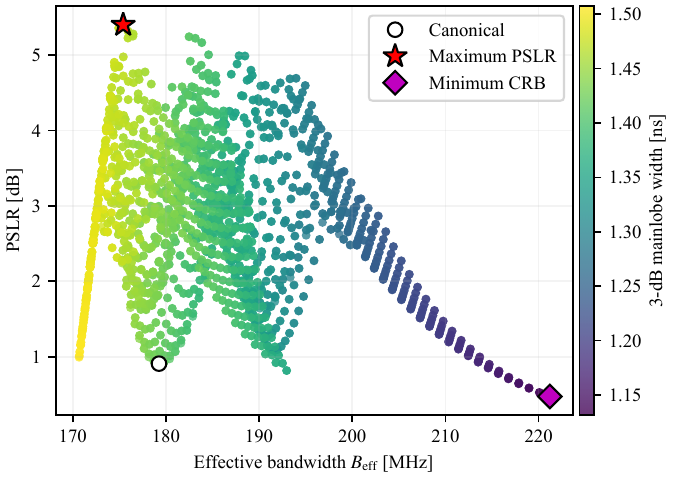}\\[-1mm]
    {\footnotesize\rev{(a) Effective bandwidth--PSLR tradeoff}}
\end{minipage}
\hfill
\begin{minipage}[t]{0.275\linewidth}
    \centering
    \includegraphics[width=\linewidth]{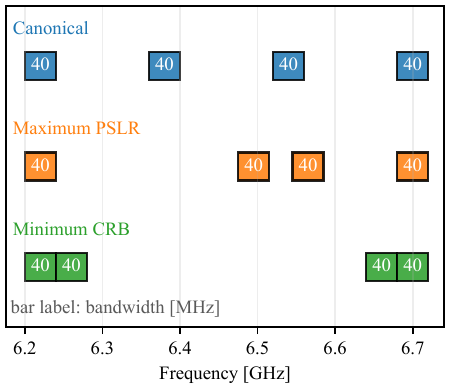}\\[-1mm]
    {\footnotesize\rev{(b) Representative layouts}}
\end{minipage}
\hfill
\begin{minipage}[t]{0.32\linewidth}
    \centering
    \includegraphics[width=\linewidth]{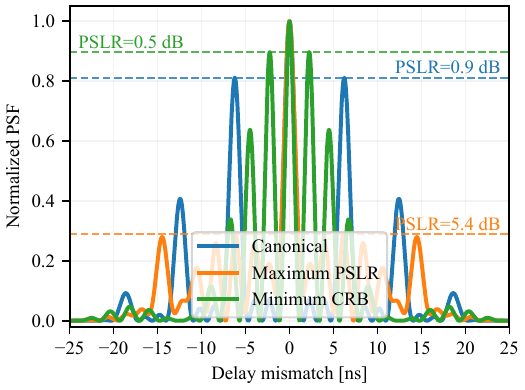}\\[-1mm]
    {\footnotesize\rev{(c) Normalized PSFs}}
\end{minipage}

\caption{\rev{Non-uniform subband layout analysis under equal spectrum resources.
(a) Tradeoff between effective bandwidth and the peak-to-sidelobe ratio (PSLR)
over feasible middle-subband placements, with color indicating the 3-dB mainlobe width.
(b) Representative canonical equal-spaced, Minimum-CRB, and Maximum-PSLR layouts.
(c) Normalized PSFs of the representative layouts.}}
\label{fig:irregular_layout_psf}
\end{figure*}

\subsection{\texorpdfstring{\polish{Canonical Equal-Spaced Subband Layout Analysis}}{Canonical Equal-Spaced Subband Layout Analysis}}
\label{sec:multi_band_selection}
\polish{
We now specialize the subband carrier frequencies \(F_t\) to the canonical
equal-spaced layout. This layout leads to a closed-form PSF that makes the
roles of subband spacing, subband width, and the number of subbands explicit.

For the canonical equal-spaced layout, the subband centers are written as
\begin{align}
F_t
&=
F_1+(t-1)f_{\mathrm{gap}},\qquad t=1,\ldots,T,
\label{eq:centers_equal_spacing}
\end{align}
where \(F_1\) is the first subband-center frequency and
\(f_{\mathrm{gap}}\) denotes the spacing between adjacent subband centers.

The coherent delay kernel in
\eqref{eq:coherent_delay_kernel_definition} gives
\begin{align}
h(\delta_\tau,\mathcal{F})
&=
\sum_{t=1}^{T}\sum_{m\in\mathcal{U}}
e^{j2\pi f_{t,m}\delta_\tau}
\notag\\
&=
\Big(\sum_{m\in\mathcal{U}}e^{j2\pi m\Delta f\,\delta_\tau}\Big)
\Big(\sum_{t=1}^{T}e^{j2\pi F_t\delta_\tau}\Big).
\label{eq:w_factored}
\end{align}
The first factor captures the intra-band contribution from the \(M_b\)
uniformly spaced measured frequencies within each subband, and the second
factor captures the inter-band contribution from the \(T\) subband centers.

The intra-band sum admits a Dirichlet-kernel form. Its squared magnitude is given by
\begin{equation}
\left|D(\delta_\tau)\right|^2
=
\left| \sum_{m\in\mathcal{U}}e^{j2\pi m\Delta f\,\delta_\tau} \right|^2
=
\left| \frac{\sin(\pi M_b\Delta f\,\delta_\tau)}
{\sin(\pi\Delta f\,\delta_\tau)} \right|^2.
\label{eq:dirichlet_def}
\end{equation}

The squared magnitude $|D(\delta_\tau)|^2$ has a dominant mainlobe around
\(\delta_\tau=0\) with first null at
\(\delta_\tau=\pm 1/(M_b\Delta f)\), which implies a mainlobe width on the
order of \(1/(M_b\Delta f)\).

For the canonical equal-spaced centers in \eqref{eq:centers_equal_spacing}, the
squared magnitude of the inter-band factor in \eqref{eq:w_factored} satisfies
\begin{equation}
\left|\sum_{t=1}^{T}e^{j2\pi F_t\delta_\tau}\right|^2
=
\left|\frac{\sin(\pi T f_{\mathrm{gap}}\,\delta_\tau)}
{\sin(\pi f_{\mathrm{gap}}\,\delta_\tau)}\right|^2.
\label{eq:comb_mag}
\end{equation}

This term exhibits a comb-like response with peak spacing
\(1/f_{\mathrm{gap}}\), and its first null around the origin occurs at
\(\delta_\tau=\pm 1/(T f_{\mathrm{gap}})\).

Combining \eqref{eq:w_factored}--\eqref{eq:comb_mag} with the PSF relation
in \eqref{eq:psf_definition} yields the PSF of the canonical
equal-spaced subband layout
\begin{align}
P(\delta_\tau,\mathcal{F})
&=
\left|h(\delta_\tau,\mathcal{F})\right|^2 \notag\\
&=
|D(\delta_\tau)|^2
\left|\frac{\sin(T\pi f_{\mathrm{gap}}\,\delta_\tau)}
{\sin(\pi f_{\mathrm{gap}}\,\delta_\tau)}\right|^2.
\label{eq:psf_multiband_final}
\end{align}

Equation~\eqref{eq:psf_multiband_final} shows that the canonical PSF is the
product of an inter-band comb term and an intra-band Dirichlet envelope. The
comb term determines the locations of the periodic peaks,
whereas the Dirichlet envelope determines how strongly these peaks are
attenuated. When \(1/f_{\mathrm{gap}}\) is smaller than the envelope mainlobe
width, which is on the order of \(1/(M_b\Delta f)\), several periodic sidelobe
peaks can remain inside the intra-band Dirichlet envelope. These periodic
sidelobe peaks at small delay offsets experience only weak attenuation and can
become competing local maxima, which leads to delay ambiguity. Increasing \(M_b\Delta f\)
narrows the envelope and attenuates sidelobe peaks farther away from
\(\delta_\tau=0\).

Fig.~\ref{fig:PSF} illustrates these tradeoffs. In Fig.~\ref{fig:PSF}a,
increasing \(T\) with fixed \(f_{\mathrm{gap}}\) and \(M_b\Delta f\) narrows the
mainlobe of the PSF while keeping the peak spacing \(1/f_{\mathrm{gap}}\)
unchanged. In Fig.~\ref{fig:PSF}b, with fixed
\(T\), increasing \(f_{\mathrm{gap}}\) also narrows the mainlobe but
decreases the peak spacing, moving the periodic sidelobe peaks closer to
\(\delta_\tau=0\) and increasing the potential delay ambiguity. In
Fig.~\ref{fig:PSF}c, increasing \(M_b\Delta f\) suppresses these sidelobe peaks
by tightening the Dirichlet envelope in \eqref{eq:psf_multiband_final}.
}

\subsection{\texorpdfstring{\rev{Non-Uniform Subband Layout Analysis}}{Non-Uniform Subband Layout Analysis}}
\label{sec:nonuniform_layout_psf}
\rev{
For a non-uniform subband layout,
the inter-subband summation generally does not reduce to the sine-ratio form in
\eqref{eq:comb_mag}. To obtain more insight into how the layout affects delay
estimation, we evaluate each feasible layout using the effective bandwidth and
the peak-to-sidelobe ratio (PSLR). The effective bandwidth is connected to the
CRB in \eqref{eq:crb_var_form_multiband}, while the PSLR measures the dominant
PSF sidelobe.

To compare different frequency layouts under the same spectrum resources, we
fix \(T=4\), \(M_b=128\), and \(\Delta f=312.5~\mathrm{kHz}\). The two edge
subband centers are fixed at \(F_1=6.22~\mathrm{GHz}\) and
\(F_4=6.7~\mathrm{GHz}\). The two middle subband centers are swept over the
feasible set
}
\rev{\begin{align}
\mathcal{A}
&=
\big\{(F_2,F_3)\mid \notag\\
&\qquad
F_{t+1}-F_t\ge M_b\Delta f,\ t=1,2,3,\notag\\
&\qquad
(F_2-F_1)/(5~\mathrm{MHz}),\
(F_3-F_1)/(5~\mathrm{MHz})\in\mathbb{Z}
\big\}.
\label{eq:nonuniform_feasible_set}
\end{align}}
\rev{
For each candidate pair \((F_2,F_3)\in\mathcal{A}\), we evaluate the
peak-to-sidelobe ratio
}
\rev{\begin{align}
\mathrm{PSLR}
&=
10\log_{10}
\frac{P(0,\mathcal{F})}
{\max_{\delta_\tau\notin\mathcal{M}_{\mathrm{main}}}
P(\delta_\tau,\mathcal{F})},
\label{eq:layout_pslr}
\end{align}}
\rev{
where \(\mathcal{M}_{\mathrm{main}}\) denotes the mainlobe region around
\(\delta_\tau=0\).

The connection to the CRB follows from the frequency-variance decomposition in
\eqref{eq:var_term_multiband_general_centers}. Define the effective bandwidth
of a layout as
}
\rev{\begin{align}
B_{\mathrm{eff}}
&=
\left[
\frac{1}{T}
\sum_{t=1}^{T}
\big(F_t-\bar F\big)^2
+
\frac{M_b^2-1}{12}(\Delta f)^2
\right]^{1/2},
\label{eq:layout_beff}
\end{align}}
\rev{
where \(\bar F=T^{-1}\sum_{t=1}^{T}F_t\) is the mean
subband-center frequency.
The first term in \eqref{eq:layout_beff} is the inter-band contribution from
the spread of the subband centers, while the second term is the intra-band
contribution from the finite subband bandwidth. All feasible layouts use the
same \(T\), \(M_b\), and \(\Delta f\), so the intra-band term is fixed and only
the spread of the subband centers changes across layouts. For fixed
\(\sigma^2\) and \(|\alpha_0|\), the denominator of
\eqref{eq:crb_var_form_multiband} is proportional to
\(TM_b B_{\mathrm{eff}}^2\). Hence, the Minimum-CRB layout is the
feasible layout that maximizes the spread of the subband centers and
equivalently \(B_{\mathrm{eff}}\). However, a larger effective
bandwidth does not necessarily imply a better sidelobe structure.

Fig.~\ref{fig:irregular_layout_psf}a summarizes all feasible frequency layouts.
As \(B_{\mathrm{eff}}\) increases, the 3-dB mainlobe width generally
becomes narrower, which indicates higher delay sensitivity. However, the layout
distribution shifts toward lower PSLR as \(B_{\mathrm{eff}}\)
becomes large. 

Fig.~\ref{fig:irregular_layout_psf}b and Fig.~\ref{fig:irregular_layout_psf}c
show the selected representative layouts and their corresponding PSFs. The
canonical layout is the equal-spaced
reference. It has a moderate \(B_{\mathrm{eff}}\).
As explained by \eqref{eq:psf_multiband_final}, its regular spacing creates
periodic sidelobe peaks, and some of these peaks remain weakly attenuated by
the Dirichlet envelope. These peaks become large dominant sidelobes and lead to
a low PSLR for this layout. The
Minimum-CRB layout moves the two middle subbands next to the two edge subbands,
so the four 40 MHz subbands form two 80 MHz contiguous frequency blocks at the
two ends of the span. This maximizes
\(B_{\mathrm{eff}}\) over the feasible set and therefore achieves
the minimum CRB. From the PSF perspective, this layout has two competing
effects. Grouping two adjacent 40 MHz subbands into each 80 MHz block narrows
the corresponding Dirichlet envelope, which helps attenuate sidelobe peaks
farther away from \(\delta_\tau=0\). At the same time, the large separation between the two blocks creates
closely spaced sidelobe peaks near \(\delta_\tau=0\). Since these peaks are
close to the mainlobe, they lie in a region where the envelope is still high and
are therefore not sufficiently suppressed. This explains why the Minimum-CRB
layout has the largest \(B_{\mathrm{eff}}\) but the lowest PSLR among the
feasible layouts in Fig.~\ref{fig:irregular_layout_psf}a. In contrast, the
Maximum-PSLR layout uses more irregular subband-center spacing to reduce the
coherent alignment of sidelobes from different subbands. This suppresses the
dominant sidelobe and gives the highest PSLR among the representative layouts.
}

\section{Multi-Band Single-Path Delay Estimation}
\label{sec:multiband_single_path_estimation}

A common LS-based delay estimator evaluates $\mathcal{C}(\tau)$ over a discrete grid of candidate delays
and selects the maximizer. This discretization incurs basis mismatch when the true delay lies between grid points.
Using a finer grid can reduce the mismatch, but at the expense of increased computational complexity.

To reduce basis mismatch without an excessively dense grid, a coarse grid search is often followed by a local
off-grid refinement step~\cite{NOMP,offgridOMP1,offgridOMP2,offgridOMP3,OffgridOMP4,interpolation1}, e.g., Newton refinement or other
interpolation-based methods. However, in the multi-band setting, $\mathcal{C}(\tau)$ exhibits a pronounced multimodal
structure, resulting in multiple closely spaced
sidelobe peaks. Consequently, local refinements become highly sensitive to the initialization provided by the coarse
search and may easily get trapped by nearby sidelobe peaks rather than converging to the global maximizer.
As discussed in Section~\ref{sec:simulation_A}, using a finer coarse-search grid can reduce this risk, but it again comes at the cost of substantially higher computational complexity.

Particle swarm optimization provides a more robust alternative in this multimodal scenario, since its
population-based search reduces the likelihood of being trapped in suboptimal local maxima while enabling
continuous-domain maximization of $\mathcal{C}(\tau)$. Nevertheless, applying PSO over the full delay range can still be
computationally expensive, typically requiring a large swarm size and many iterations for reliable convergence.
\polish{
To balance robustness and complexity, we adopt a two-stage PSO (TSPSO) delay
estimator. In the first stage, one contiguous reference subband is used to
locate the mainlobe of the intra-band Dirichlet envelope. A low-complexity grid
search on the reference-subband objective gives a coarse estimate
\(\hat{\tau}_{\mathrm{coarse}}\), which sets the center of the local interval
\(\mathcal{I}_{\mathrm{ref}}\). This interval substantially reduces the PSO
search range, so fewer particles and iterations are needed for the subsequent
continuous-domain refinement. In the second stage, PSO is initialized inside
\(\mathcal{I}_{\mathrm{ref}}\) and maximizes the full multi-band objective
\(\mathcal{C}(\tau)\), thereby exploiting the multi-band aperture for
high-accuracy delay estimation.

}

\subsection{\texorpdfstring{\polish{First-Stage Coarse Delay Estimation}}{First-Stage Coarse Delay Estimation}}
\label{sec:coarse_sb}

\polish{
The subband indexed by \(t=1\) is used as the reference subband for the
first-stage coarse search. Its measured frequencies \(f_{1,m}\), with
\(m\in\mathcal{U}\), are collectively denoted by
\(\mathcal{F}_{\mathrm{ref}}\). The vector
\(\mathbf{y}_{\mathrm{ref}}\) contains the entries of \(\mathbf{y}\) measured
on \(\mathcal{F}_{\mathrm{ref}}\), and
\(\mathbf{a}_{\mathrm{ref}}(\tau)=
\big[\,e^{-j2\pi f_{1,m}\tau}\,\big]_{m\in\mathcal{U}}^{\!\top}\).

By the same LS concentration as in \eqref{eq:objective_def_generic}, the
reference-subband objective is
\begin{align}
\mathcal{C}_{\mathrm{ref}}(\tau)
&=
\frac{\big|\mathbf{a}_{\mathrm{ref}}(\tau)^{H}\mathbf{y}_{\mathrm{ref}}\big|^2}
{\left\|\mathbf{a}_{\mathrm{ref}}(\tau)\right\|^{2}} .
\label{eq:objective_ref_def}
\end{align}
Restricting the single-path model in \eqref{eq:vec_model_single} to the
reference subband and taking expectation over the noise gives
\begin{align}
\mathbb{E}\!\left[
\left|\mathbf{a}_{\mathrm{ref}}(\tau)^{H}\mathbf{y}_{\mathrm{ref}}\right|^2
\right]
&=
|\alpha_0|^2\left|D(\tau-\tau_0)\right|^2+\sigma^2M_b,
\notag\\
\mathbb{E}\big[\mathcal{C}_{\mathrm{ref}}(\tau)\big]
&=
\frac{|\alpha_0|^2}{M_b}\left|D(\tau-\tau_0)\right|^2+\sigma^2.
\label{eq:ref_single_path_relation}
\end{align}
where \(D(\cdot)\) follows the Dirichlet-kernel form in
\eqref{eq:dirichlet_def}. This structure motivates the reference-subband grid
search. Let \([\tau_{\min},\tau_{\max}]\) denote the coarse-search interval.
Since the reference-subband power response
$|D(\tau-\tau_0)|^2$ is periodic with period $1/\Delta f$,
we restrict the coarse-search interval to no more than half
a period, i.e.,
$\tau_{\max}-\tau_{\min}\leq 1/(2\Delta f)$.
We set the coarse-grid spacing to
$\Delta_{\mathrm{g}}=1/(M_b\Delta f)$ and evaluate
$\mathcal{C}_{\mathrm{ref}}(\tau)$ over the following discrete
candidate set 
\begin{align}
\mathcal{G}
=
\left\{
\tau_{\min}+n\Delta_{\mathrm{g}}
\;\middle|\;
n=0,\ldots,
\left\lfloor
\frac{\tau_{\max}-\tau_{\min}}
{\Delta_{\mathrm{g}}}
\right\rfloor
\right\}
\cup\{\tau_{\max}\}.
\label{eq:coarse_ref_grid}
\end{align}
and define
\begin{align}
\hat{\tau}_{\mathrm{coarse}}
&=
\arg\max_{\tau\in\mathcal{G}}
\mathcal{C}_{\mathrm{ref}}(\tau).
\label{eq:coarse_ref_estimate}
\end{align}
Appendix~\ref{app:dirichlet_props} shows that
\(|D(\tau-\tau_0)|^2\) has a unique dominant peak at
\(\tau=\tau_0\) and that \( |D(\tau-\tau_0)|<|D(\Delta_{\mathrm{g}}/2)| \) for all
\(|\tau-\tau_0|>\Delta_{\mathrm{g}}/2\). In the noiseless single-path setting, this
property gives a grid-quantization bound for the reference-subband coarse
estimate. Since \(\mathcal{G}\) contains a point within \(\Delta_{\mathrm{g}}/2\) of
\(\tau_0\) when \(\tau_0\in[\tau_{\min},\tau_{\max}]\), the resulting coarse
estimate satisfies
\begin{align}
\big|\hat{\tau}_{\mathrm{coarse}}-\tau_0\big|
&\le
\frac{\Delta_{\mathrm{g}}}{2}
=
\frac{1}{2M_b\Delta f}.
\label{eq:grid_quant_bound_ref}
\end{align}
In noisy or multipath channels,
noise and unresolved paths perturb \(\mathcal{C}_{\mathrm{ref}}(\tau)\), so
the bound in \eqref{eq:grid_quant_bound_ref} is no longer deterministic. We
therefore use the noiseless bound as a reference scale and introduce a margin
factor \(\gamma\) to define the refinement interval as
\begin{align}
\mathcal{I}_{\mathrm{ref}}
&=
\Big[\hat{\tau}_{\mathrm{coarse}}-r,\,
     \hat{\tau}_{\mathrm{coarse}}+r\Big].
\label{eq:search_space}
\end{align}
where the radius is
\begin{align}
r
&\triangleq
\gamma\,\frac{\Delta_{\mathrm{g}}}{2}
=
\frac{\gamma}{2M_b\Delta f},
\qquad \gamma \ge 1.
\label{eq:search_radius}
\end{align}

Successful refinement requires the target path delay to lie within
\(\mathcal{I}_{\mathrm{ref}}\). The probability of this event under noise and
multipath is evaluated empirically in
Section~\ref{sec:performance_comparison}.
}

\subsection{Second-Stage PSO Refinement}

\begin{figure}[t]
\centering
\includegraphics[width=0.7\linewidth]{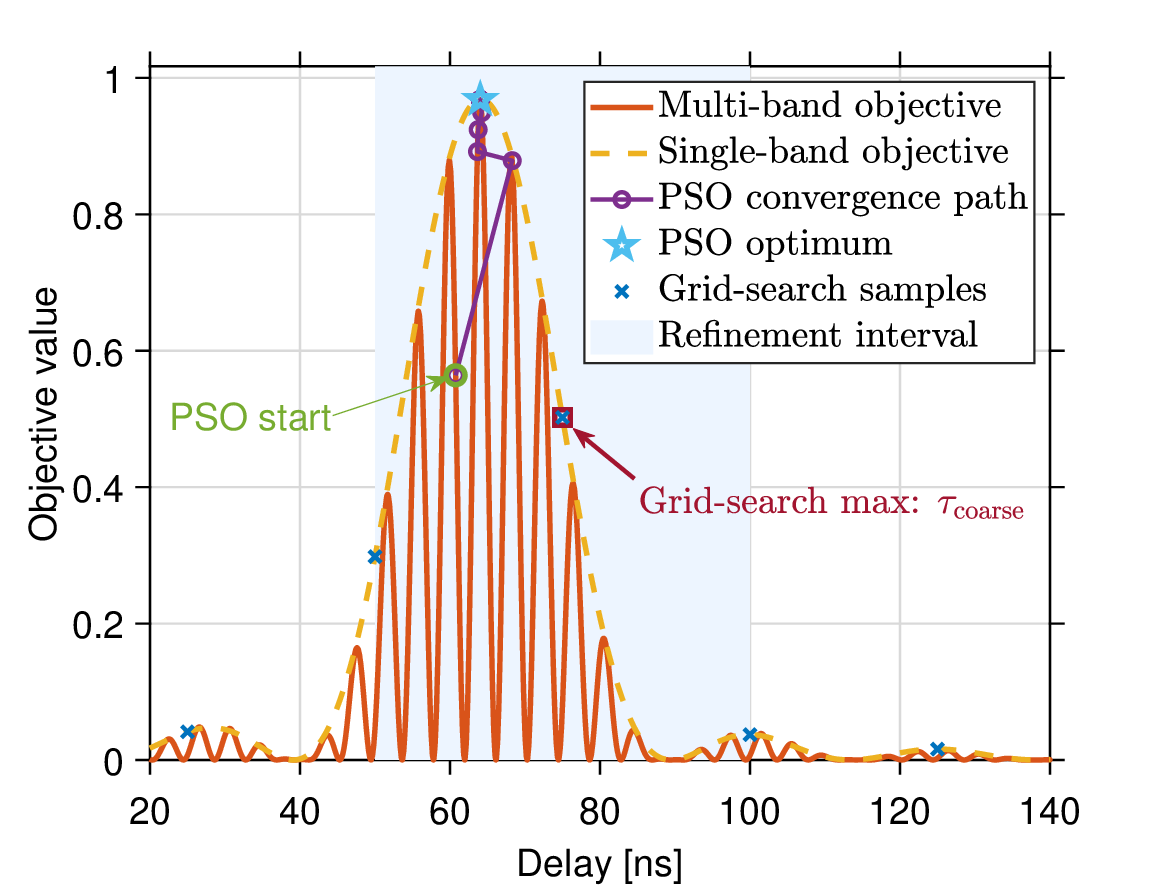}
\caption{TSPSO delay estimation for a single target at delay $63$ ns, $f_{gap}=240\,\mathrm{MHz}$, $T=2$, $M_b\Delta f=40\,\mathrm{MHz}$, $\gamma = 2$, and $\Delta_{\mathrm{g}} = 25$~ns.}
\label{fig:pso_obj}
\end{figure}

PSO is a population-based stochastic optimization method in which a set of $N_p$ particles is evolved, each representing a candidate delay $\tau$. At each iteration, particle positions and velocities are updated based on their personal best positions and the best position found by the swarm, which allows information to be shared among particles and provides a trade-off between exploration and exploitation~\cite{PSOINTRO}.  PSO requires only objective-function evaluations and does not rely on convexity or differentiability.
Its computational cost scales linearly with the number of particles $N_p$ and iterations $Q$, making it well suited for one-dimensional and highly multimodal objectives such as $\mathcal{C}(\tau)$.

In our implementation, the $N_p$ particles are initialized with delays drawn uniformly from $\mathcal{I}_{\mathrm{ref}}$
and with small random velocities, and are then iteratively updated to increase $\mathcal{C}(\tau)$.
Although PSO is heuristic and does not guarantee global optimality for a finite swarm, it has been reported
to perform effectively in multimodal optimization problems~\cite{CaoCLPSO}. In our two-stage design, the coarse
\polish{reference-subband estimate} confines the refinement to the interval $\mathcal{I}_{\mathrm{ref}}$, which substantially reduces
the search space and lowers the likelihood of convergence to distant local maxima. As a result, the proposed
estimator achieves near-CRB performance in the considered scenarios (Section~\ref{sec:simulation_A}).

Formally, at iteration $t$ the $i$th particle maintains a delay estimate $\tau_i^{(q)}$ and a velocity $v_i^{(q)}$, which are updated as
\begin{align}
v_i^{(q+1)} &= w^{(q)} v_i^{(q)}
+ c_1\,r_{1,i}^{(q)}\big(p_i^{(q)} - \tau_i^{(q)}\big)
+ c_2\,r_{2,i}^{(q)}\big(g^{(q)} - \tau_i^{(q)}\big),
\label{eq:pso_v_update} \\[2pt]
\tau_i^{(q+1)} &= 
\mathrm{proj}_{\mathcal{I}_{\mathrm{ref}}}
\Bigl(\tau_i^{(q)} + v_i^{(q+1)}\Bigr),
\label{eq:pso_tau_update}
\end{align}
where $p_i^{(q)}$ denotes the best delay found by particle $i$ up to iteration $q$, and $g^{(q)}$ is the best delay found among all particles. The random coefficients $r_{1,i}^{(q)}$ and $r_{2,i}^{(q)}$ are independent and uniformly distributed in $(0,1)$ and introduce stochastic exploration around the personal and global best positions. The constants $c_1$ and $c_2$ control the relative influence of individual memory and swarm cooperation, while the inertia weight $w^{(q)}$ is typically decreased linearly from $w_s$ to $w_e$ to gradually shift the swarm from global exploration to local refinement. The projection operator $\mathrm{proj}_{\mathcal{I}_{\mathrm{ref}}}(\cdot)$ ensures that all particles remain inside the refinement interval $\mathcal{I}_{\mathrm{ref}}$. The algorithm terminates when either a maximum number of iterations $Q$ is reached or the improvement of $\mathcal{C}(\tau)$ around the current global best becomes negligible. The resulting TSPSO procedure is summarized in Algorithm~\ref{AG:PSO_DE}.

Fig.~\ref{fig:pso_obj} illustrates the proposed TSPSO for a single target at
$\tau_0=63$~ns with $N_p=15$. The first stage provides $\hat{\tau}_{\mathrm{coarse}}$ and defines the
refinement interval $\mathcal{I}_{\mathrm{ref}}$, as highlighted in the figure. The second stage initializes PSO within
$\mathcal{I}_{\mathrm{ref}}$ and iteratively updates the particles to maximize the multi-band objective, converging to a
high-quality solution near the dominant peak.

\begin{algorithm}[t]
\caption{Two-Stage PSO Delay Estimation}
\label{AG:PSO_DE}
\KwIn{\polish{\(\mathbf{y}\), \(\mathcal{F}\), \(\mathcal{F}_{\mathrm{ref}}\), \(\gamma = 2\), and}
\\  \polish{PSO parameters \(N_p,Q,w_s,w_e,c_1,c_2,\varepsilon\).}}
\KwOut{Estimated delay $\hat{\tau}$.}

\polish{Construct \(\mathcal{C}_{\mathrm{ref}}(\tau)\) from the entries of \(\mathbf{y}\) on \(\mathcal{F}_{\mathrm{ref}}\) using \eqref{eq:objective_ref_def}}\\
\polish{Construct \(\mathcal{C}(\tau)\) from \(\mathbf{y}\) and \(\mathcal{F}\) using \eqref{eq:objective_def_generic}}\\
\medskip
\textbf{\polish{Stage 1. Coarse reference-subband search}}\\
\polish{Compute $\hat{\tau}_{\mathrm{coarse}}$ by \eqref{eq:coarse_ref_estimate}}\\
\polish{Form the refinement interval $\mathcal{I}_{\mathrm{ref}}$ by
\eqref{eq:search_space} and \eqref{eq:search_radius}} \\
\medskip
\textbf{\polish{Stage 2. PSO multi-band refinement on $\mathcal{I}_{\mathrm{ref}}$}}\\
\textit{Initialization}\\
\For{$i=1,\ldots,N_p$}{
    draw $\tau_i^{(0)}\in\mathcal{I}_{\mathrm{ref}}$ and a small random $v_i^{(0)}$\\
    set $p_i^{(0)}=\tau_i^{(0)}$
}
\polish{Set \(g^{(0)}\leftarrow
\arg\max_{\tau\in\{\tau_1^{(0)},\ldots,\tau_{N_p}^{(0)}\}}
\mathcal{C}(\tau)\)}\\

\textit{Iterations}\\
\For{$q=0,\ldots,Q-1$}{
    update inertia weight
    $w^{(q)}=w_s-\tfrac{q}{Q-1}(w_s-w_e)$\\
    \For{$i=1,\ldots,N_p$}{
        Update velocity $v_i^{(q+1)}$ using \eqref{eq:pso_v_update}\\
        Update position $\tau_i^{(q+1)}$ using \eqref{eq:pso_tau_update}\\[1pt]
        \If{\(\mathcal{C}(\tau_i^{(q+1)})>
        \mathcal{C}(p_i^{(q)})\)}{
            set \(p_i^{(q+1)}=\tau_i^{(q+1)}\)
        }
        \Else{
            set \(p_i^{(q+1)}=p_i^{(q)}\)
        }
    }
    \polish{Set \(g^{(q+1)}\leftarrow
    \arg\max_{\tau\in\{p_1^{(q+1)},\ldots,p_{N_p}^{(q+1)}\}}
    \mathcal{C}(\tau)\)}\\
    \If{$|\mathcal{C}(g^{(q+1)})
          -\mathcal{C}(g^{(q)})|<\varepsilon$}{
        $\hat{\tau}=g^{(q+1)}$ \quad \textbf{break}
    }
}
\If{no early termination}{$\hat{\tau}=g^{(Q)}$.}
\end{algorithm}

\section{Multi-band Multipath Delay Estimation}
\label{sec:dual_band_multipath}

\subsection{\texorpdfstring{\polish{Coherent Multi-Band Multipath Delay Estimation}}{Coherent Multi-Band Multipath Delay Estimation}}
\label{sec:coherent_multipath_estimation}

\polish{We now extend the single-path model in \eqref{eq:vec_model_single} to the multipath case. The corresponding observation model is
\begin{equation}
\mathbf{y}
= \mathbf{S}\boldsymbol{\alpha}+\mathbf{v},
\quad
\mathbf{S}
=\big[\mathbf{a}(\tau_1),\ldots,
      \mathbf{a}(\tau_L)\big]\in\mathbb{C}^{M\times L},
\label{eq:model_multi}
\end{equation}}
where $\boldsymbol{\alpha}=[\alpha_1,\ldots,\alpha_L]^{\top}$ collects
the path gains. The joint estimation of the delay vector
$\boldsymbol{\tau}=[\tau_1,\ldots,\tau_L]^{\top}$ and gains
$\boldsymbol{\alpha}$ can be formulated as the LS problem
\begin{equation}
R(\boldsymbol{\alpha},\boldsymbol{\tau})
= \big\|\mathbf{y}-\mathbf{S}\boldsymbol{\alpha}\big\|^2.
\label{eq:multi_ls}
\end{equation}
Directly minimizing \eqref{eq:multi_ls} with respect to
$\boldsymbol{\tau}$ leads to a high-dimensional, nonconvex optimization
problem whose complexity quickly becomes prohibitive as $L$ grows.

In practice, wideband channels are typically sparse in the delay domain,
and one often resorts to greedy sparse-recovery or relaxation-based
parametric estimators, such as OMP and RELAX, which are usually
implemented on a discrete set of candidate delays. As discussed for the
single-path case in Section~\ref{sec:multiband_single_path_estimation}, such
discrete implementations are affected by a basis-mismatch versus
complexity tradeoff. The main idea here is to replace this discrete
search by the proposed TSPSO estimator. \polish{At each path-detection
step, TSPSO is applied to the current multi-band residual. Its first stage uses
the reference-subband part of that residual only to construct the refinement
interval, and its second stage maximizes the full multi-band residual objective
over this interval.} Embedding TSPSO into the OMP and RELAX frameworks yields
the proposed PSO-OMP and PSO-RELAX algorithms.
\paragraph{PSO-OMP}
We first instantiate the above idea within an OMP framework. Classical
OMP builds up the multipath model iteratively. At each iteration it
selects a new delay, augments the dictionary with the corresponding
steering vector, and then refits all path gains by least squares \cite{hisac}.  After $p$ paths have been detected, the dictionary is given by \(\mathbf{S}^{(p)}=\big[\mathbf{a}(\hat{\tau}_1),\ldots,\mathbf{a}(\hat{\tau}_p)\big].\)  Given $\mathbf{S}^{(p)}$, the complex gains are updated via LS as
\begin{equation}
\hat{\boldsymbol{\alpha}}^{(p)}
=\big((\mathbf{S}^{(p)})^{H}\mathbf{S}^{(p)}\big)^{-1}(\mathbf{S}^{(p)})^{H}\mathbf{y},
\label{eq:refit_amp}
\end{equation}
and the associated residual is
\begin{equation}
\mathbf{r}^{(p)}=\mathbf{y}-\mathbf{S}^{(p)}\hat{\boldsymbol{\alpha}}^{(p)}.
\label{eq:refit_residual}
\end{equation}

At iteration $(p{+}1)$, the objective function for detecting the next path
is defined as
\begin{equation}\label{eq:omp_objective}
\mathcal{C}_{p+1}(\tau)
=\frac{\big|\mathbf{a}(\tau)^{H}\mathbf{r}^{(p)}\big|^2}{\left\|\mathbf{a}(\tau)\right\|^{2}},
\end{equation}
and the next delay estimate is obtained by \( \hat{\tau}_{p+1}
=\arg\max_{\tau}\,\mathcal{C}_{p+1}(\tau),\) where \polish{TSPSO carries
out the continuous-domain maximization of this multi-band residual objective.}
 After obtaining $\hat{\tau}_{p+1}$, the dictionary
is augmented with $\mathbf{a}(\hat{\tau}_{p+1})$, followed by LS refitting and residual update via
\eqref{eq:refit_amp}--\eqref{eq:refit_residual}. The overall procedure is summarized in Algorithm~\ref{alg:pso_omp}.

\begin{algorithm}[t]
\caption{PSO-OMP Multipath Delay Estimation}
\label{alg:pso_omp}
\KwIn{$\mathbf{y}$ and maximum number of paths $L$.}
\KwOut{Estimated delays $\{\hat{\tau}_\ell\}_{\ell=1}^{L}$ and gains $\{\hat{\alpha}_\ell\}_{\ell=1}^{L}$.}

\textbf{Initialization}
$\mathbf{r}^{(0)}\leftarrow\mathbf{y}$,\quad
$\mathbf{S}^{(0)}\leftarrow[\ ]$.

\textbf{OMP Loop}
\For{$p=0$ \KwTo $L-1$}{
  \emph{(1) Path detection (TSPSO)}\\
  \hspace*{1em}\polish{Run Algorithm~\ref{AG:PSO_DE} with \(\mathbf{y}\leftarrow\mathbf{r}^{(p)}\)}\\
  \hspace*{1em}\polish{to obtain \(\hat{\tau}_{p+1}\)}\\[2pt]
  \emph{(2) Dictionary update}\\
    \hspace*{1em}$\mathbf{S}^{(p+1)} \leftarrow [\mathbf{S}^{(p)}\ \mathbf{a}(\hat{\tau}_{p+1})]$\\[3pt]
  \emph{(3) Residual update}\\
  \hspace*{1em} Update $\hat{\boldsymbol{\alpha}}^{(p+1)}$ and $\mathbf{r}^{(p+1)}$
  using \eqref{eq:refit_amp}--\eqref{eq:refit_residual}.\\
}
\end{algorithm}
\paragraph{PSO-RELAX}
PSO-RELAX shares the same path detection and dictionary update steps as PSO-OMP, 
but adds a cyclic refinement stage after each new path detection~\cite{Relax2,Relax1,dualband}. 
This stage mitigates inter-path interference by sequentially refining the detected paths using Newton refinement~\cite{NOMP}.

After detecting the $(p{+}1)$th path, it performs $R_c$ refinement cycles. In each cycle, the paths are updated sequentially.
For a given index $\ell$, we form the exclusion residual
\begin{equation}
\mathbf{J}_{\ell}
= \mathbf{y} - \sum_{i=1,\,i\neq \ell}^{p+1}\hat{\alpha}_i\,\mathbf{a}(\hat{\tau}_i),
\label{eq:relax_ex_res}
\end{equation}
which removes the contribution of all paths except the $\ell$th one. Conditioned on $\mathbf{J}_{\ell}$, we define the corresponding objective
\begin{equation}
\mathcal{C}_{\mathbf{J}}(\hat{\alpha}_\ell,\hat{\tau}_\ell)
=2\Re\left\{\mathbf{J}_{\ell}^{H}\hat{\alpha}_\ell\mathbf{a}(\hat{\tau}_\ell)\right\}
-|\hat{\alpha}_\ell|^{2}\left\|\mathbf{a}(\hat{\tau}_\ell)\right\|^{2},
\label{eq:f_nomptype}
\end{equation}
where $(\hat{\alpha}_{\ell},\hat{\tau}_{\ell})$ denote the current estimate. The refinement of the $\ell$th path delay is then performed by a Newton update
\begin{equation}
\hat{\tau}_{\ell}'=\hat{\tau}_{\ell}-\frac{\dot{\mathcal{C}_{\mathbf{J}}}(\hat{\alpha}_{\ell},\hat{\tau}_{\ell})}
{\ddot{\mathcal{C}_{\mathbf{J}}}(\hat{\alpha}_{\ell},\hat{\tau}_{\ell})}.
\label{eq:newton_relax_update}
\end{equation}
where
\begin{align}
\dot{\mathcal{C}_{\mathbf{J}}}(\hat{\alpha}_{\ell},\hat{\tau}_{\ell})
&=2 \Re\left\{\big(\mathbf{J}_{\ell}-\hat{\alpha}_{\ell}\mathbf{a}(\hat{\tau}_{\ell})\big)^{H}\,
\hat{\alpha}_{\ell}\left(\frac{d \mathbf{a}(\hat{\tau}_{\ell})}{d\hat{\tau}_{\ell}}\right)\right\}, \label{eq:newton_f}\\ \notag
\ddot{\mathcal{C}_{\mathbf{J}}}(\hat{\alpha}_{\ell},\hat{\tau}_{\ell})
&=2 \Re\left\{\big(\mathbf{J}_{\ell}-\hat{\alpha}_{\ell}\mathbf{a}(\hat{\tau}_{\ell})\big)^{H}\,
\hat{\alpha}_{\ell}\left(\frac{d^{2}\mathbf{a}(\hat{\tau}_{\ell})}{d\hat{\tau}_{\ell}^{2}}\right)\right\}\\ \notag
& -2|\hat{\alpha}_{\ell}|^{2}\left\|\frac{d \mathbf{a}(\hat{\tau}_{\ell})}{d\hat{\tau}_{\ell}}\right\|^{2}. 
\end{align}

Then, the gain is updated in closed form as \(\hat{\alpha}_{\ell}'=(\mathbf{a}(\hat{\tau}_{\ell}')^{H}\mathbf{J}_{\ell})/\left\|\mathbf{a}(\hat{\tau}_\ell')\right\|^{2}. \)
After one cycle over $\ell=1,\dots,p{+}1$, we recompute all $(p{+}1)$ gains
by an LS refit using the refined delays, and update the residual via
\eqref{eq:refit_amp}--\eqref{eq:refit_residual}. This LS refitting step
ensures that the gains and residual are jointly optimal for the current
delay estimates. The complete PSO-RELAX procedure is summarized in
Algorithm~\ref{alg:pso_relax}.

\begin{algorithm}[t]
\caption{PSO-RELAX Multipath Delay Estimation}
\label{alg:pso_relax}
\KwIn{$\mathbf{y}$, maximum number of paths $L$, and maximum relaxation cycles $R_c$.}
\KwOut{Estimated delays $\{\hat{\tau}_\ell\}_{\ell=1}^{L}$ and gains $\{\hat{\alpha}_\ell\}_{\ell=1}^{L}$.}

\textbf{Initialization}
$\mathbf{r}^{(0)} \leftarrow \mathbf{y}$,\quad
$\mathbf{S}^{(0)} \leftarrow [\ ]$.

\textbf{OMP outer loop}\\
\For{$p=0$ \KwTo $L-1$}{
    \emph{(1) Path detection (TSPSO)}\\
    \hspace*{1em}\polish{Run Algorithm~\ref{AG:PSO_DE} with \(\mathbf{y}\leftarrow\mathbf{r}^{(p)}\)}\\
    \hspace*{1em}\polish{to obtain \(\hat{\tau}_{p+1}\)}\\[2pt]

    \emph{(2) Dictionary update}\\
    \hspace*{1em}$\mathbf{S}^{(p+1)} \leftarrow [\mathbf{S}^{(p)}\ \mathbf{a}(\hat{\tau}_{p+1})]$\\[3pt]

    \emph{(3) Residual update}\\

        \hspace*{1em}Compute $\hat{\boldsymbol{\alpha}}^{(p+1)}$ and $\mathbf{r}^{(p+1)}$ via \eqref{eq:refit_amp}--\eqref{eq:refit_residual}\\
    \emph{(4) Relaxation cycles}\\
    \For{$j=0$ \KwTo $R_c-1$}{
        \For{$\ell=1$ \KwTo $p{+}1$}{
            Construct the exclusion residual $\mathbf{J}_{\ell}$ using \eqref{eq:relax_ex_res}\\
            Update $\hat{\tau}_\ell'$ and $\hat{\alpha}_\ell'$ via \eqref{eq:relax_ex_res}--\eqref{eq:newton_f}\\
            Replace the $\ell$th column of $\mathbf{S}^{(p+1)}$ by $\mathbf{a}(\hat{\tau}_\ell')$.
        }
        Recompute $\hat{\boldsymbol{\alpha}}^{(p+1)}$ and $\mathbf{r}^{(p+1)}$ via \eqref{eq:refit_amp}--\eqref{eq:refit_residual}.\\
    }
}
\end{algorithm}

\paragraph{\texorpdfstring{\rev{Path-Order Selection}}{Path-Order Selection}}
\rev{
When the number of paths is unknown, \(L\) in Algorithms~\ref{alg:pso_omp}
and~\ref{alg:pso_relax} is treated as an upper bound \(L_{\max}\).
Both sequential estimators can be run up to \(L_{\max}\), producing candidate
models with orders \(K=0,\ldots,L_{\max}\). The order is selected using the
Bayesian information criterion (BIC)~\cite{Ding2018ModelSelection}.
For the \(K\)-path candidate, after LS refitting of the path gains, the residual
energy and BIC score are
}
\rev{\begin{align}
\mathrm{RSS}(K)
&=
\left\|
\mathbf{y}
-
\mathbf{S}(\hat{\boldsymbol{\tau}}_K)
\hat{\boldsymbol{\alpha}}_K
\right\|_2^2,
\notag\\
\mathrm{BIC}(K)
&=
2M\log\!\left(\frac{\mathrm{RSS}(K)}{M}\right)
+
(3K+1)\log M .
\label{eq:bic_score}
\end{align}}
\rev{
where \(3K+1\) counts the \(K\) delays, the real and imaginary parts of the
\(K\) complex gains, and the noise variance. The selected order is
}
\rev{\begin{align}
\hat{L}
&=
\arg\min_{0\le K\le L_{\max}}
\mathrm{BIC}(K).
\label{eq:bic_order}
\end{align}}
\rev{
The selected candidate model provides the final delay and gain estimates.
}

\paragraph{Computational Complexity Analysis}
We summarize the dominant arithmetic complexity of TSPSO, PSO-OMP, and PSO-RELAX in terms of objective-function
evaluations. We assume direct computation of inner products, so evaluating the \polish{reference-subband objective}
$\mathcal{C}_{\mathrm{ref}}(\tau)$ costs $O(M_b)$ operations and evaluating the multi-band objective
$\mathcal{C}(\tau)$ costs $O(M)$ operations.

\textbf{One TSPSO call.} Stage~1 performs a coarse grid search over $N_g=|\mathcal{G}|$ delay candidates, which
requires $N_g$ evaluations of $\mathcal{C}_{\mathrm{ref}}(\tau)$ and therefore costs $O(M_b N_g)$. Stage~2 runs
PSO with $N_p$ particles for $Q$ iterations, requiring $N_pQ$ evaluations of the multi-band objective
$\mathcal{C}(\tau)$ and therefore costing $O(M N_p Q)$. Hence, one TSPSO invocation has overall complexity
$O(M_b N_g + M N_p Q)$.

\textbf{PSO-OMP.} At each OMP iteration, TSPSO is invoked once using the current residual to identify the next
delay component. For $L$ detected paths, PSO-OMP makes $L$ TSPSO calls, resulting in complexity
$O\!\big(L(M_b N_g + M N_p Q)\big)$.

\textbf{PSO-RELAX.} PSO-RELAX performs $L$ outer OMP iterations and thus invokes TSPSO $L$ times for path detection, 
resulting in a complexity of $O\big(L(M_b N_g + M N_p Q)\big)$. After detecting the $(p{+}1)$th path, it further carries out
$R_c$ cyclic refinement cycles, where the $(p{+}1)$ detected paths are updated sequentially. Since each Newton refinement costs
$O(M)$, the total refinement complexity scales as $O(M R_c L^2)$. Therefore, the overall complexity of PSO-RELAX is \(O\Big(L(M_b N_g + M N_p Q) + M R_c L^2\Big). \)

\subsection{\texorpdfstring{\rev{Impact of Inter-Band Non-Coherence}}{Impact of Inter-Band Non-Coherence}}
\label{sec:interband_noncoherence}
\rev{

The coherent multi-band model in \eqref{eq:model_multi} assumes that the selected
subbands share a common phase reference, allowing their CFR
measurements to be coherently combined. In practice, such inter-band
coherence can be established through a handshaking procedure, where the
forward and reverse CFR measurements are combined to cancel Tx/Rx
phase distortions under channel reciprocity~\cite{subspace2,Decimeter}.
Alternatively, a common anchor path observed across the selected subbands can be used to estimate and compensate the inter-band phase distortions, thereby aligning the subband CFRs~\cite{hisac,jialunradar}.

When such compensation is unavailable or imperfect, residual synchronization
errors lead to inter-band non-coherence. To evaluate its impact, we represent
the remaining distortions by effective subband-dependent parameters. The phase
offset \(\phi_t\) captures constant phase rotations in subband
\(t\), including the random phase introduced by phase-locked loop (PLL)
relocking and the residual carrier-frequency-offset (CFO) phase accumulated
over the considered snapshot~\cite{hisac}. The equivalent timing offset
\(\epsilon_t\) captures frequency-linear phase terms associated with symbol
timing offset (STO), sampling frequency offset (SFO), or packet-boundary timing
uncertainty~\cite{Splicer}.  The reference subband defined in
Section~\ref{sec:coarse_sb} serves as the common timing and phase reference,
with \(\phi_1=0\) and \(\epsilon_1=0\). The resulting phase distortion is
represented by
}
\rev{\begin{align}
\mathbf{G}_t
&=
\operatorname{diag}
\left(
\left\{
e^{-j[\phi_t+2\pi m\Delta f\,\epsilon_t]}
\right\}_{m\in\mathcal{U}}
\right),
\notag\\
\mathbf{G}(\boldsymbol{\phi},\boldsymbol{\epsilon})
&=
\operatorname{blkdiag}
\big(\mathbf{G}_1,\ldots,\mathbf{G}_T\big),
\label{eq:subband_nonideality}
\end{align}}
\rev{
where \(\boldsymbol{\phi}=[\phi_1,\ldots,\phi_T]^T\) and
\(\boldsymbol{\epsilon}=[\epsilon_1,\ldots,\epsilon_T]^T\). Under these
relative offsets, the coherent model in \eqref{eq:model_multi} becomes the
non-coherent observation model
}
\rev{\begin{align}
\mathbf{y}_{\mathrm{nc}}
&=
\mathbf{G}(\boldsymbol{\phi},\boldsymbol{\epsilon})
\mathbf{S}\boldsymbol{\alpha}
+\mathbf{v}.
\label{eq:phase_offset_model}
\end{align}}
\rev{
When the objective in \eqref{eq:omp_objective} is evaluated on
\(\mathbf{y}_{\mathrm{nc}}\) at the first path-detection step, the
relevant inner product reduces to \(\mathbf{a}(\tau)^H\mathbf{y}_{\mathrm{nc}}\), which can be written as
}
\rev{\begin{align}
\mathbf{a}(\tau)^H\mathbf{y}_{\mathrm{nc}}
&=
\sum_{\ell=1}^{L}
\alpha_\ell
\mathbf{a}(\tau)^H
\mathbf{G}(\boldsymbol{\phi},\boldsymbol{\epsilon})
\mathbf{a}(\tau_\ell)
+\mathbf{a}(\tau)^H\mathbf{v}
\notag\\
&=
\sum_{\ell=1}^{L}\alpha_\ell
\sum_{t=1}^{T}
e^{-j\phi_t}
e^{j2\pi F_t(\tau-\tau_\ell)}
\notag\\
&\qquad\times
D(\tau-\tau_\ell-\epsilon_t)
+\mathbf{a}(\tau)^H\mathbf{v},
\label{eq:noncoherent_delay_projection}
\end{align}}
\rev{
where \(D(\cdot)\) follows the Dirichlet-kernel form in
\eqref{eq:dirichlet_def}. To examine the two offset mechanisms separately, we
first retain only the phase offsets by setting all timing offsets \(\epsilon_t\) to zero. At \(\tau=\tau_\ell\), the contribution of path \(\ell\) becomes
\(\alpha_\ell M_b\sum_{t=1}^{T}e^{-j\phi_t}\). Under phase alignment, all
terms \(e^{-j\phi_t}\) point in the same direction, giving a response magnitude
of \(|\alpha_\ell|M_bT\). When the phase offsets differ, these terms point in
different directions, so the magnitude of the multi-band response is generally reduced. For \(\tau\neq\tau_\ell\), these phase rotations alter the coherent
summation of the delay-dependent subband terms in
\eqref{eq:noncoherent_delay_projection}, thereby distorting the
sidelobe pattern of the multi-band response.

The timing-offset effect is then examined by setting all phase offsets
\(\phi_t\) to zero. The contribution from subband \(t\) contains the delay
kernel \(D(\tau-\tau_\ell-\epsilon_t)\), whose magnitude is maximized at
\(\tau=\tau_\ell+\epsilon_t\). Therefore, non-identical timing offsets shift the
subband responses to different delay locations. The subband responses then no
longer add coherently at a common delay, which reduces the main peak and can
broaden or distort the overall multi-band response. 

Therefore, phase offsets rotate the subband contributions and weaken
coherent addition across subbands, whereas timing offsets shift the
per-subband delay responses by different amounts and misalign their main
peaks. Both effects reduce the dominance of the response at the true
path delay and can make nearby sidelobe peaks more competitive, thereby
introducing delay bias and ambiguity. The resulting performance
degradation is quantified in Section~\ref{sec:phase_offset_sim}.

}

\section{\texorpdfstring{\polish{Single-Path Cram\'er--Rao Bound for Multi-Band Configurations}}{Single-Path Cramer-Rao Bound for Multi-Band Configurations}}
\label{sec:crb}

\polish{ 
In this section, we derive the CRB for delay estimation under the coherent
single-path model in \eqref{eq:vec_model_single}. The CRB lower-bounds the
variance of any unbiased delay estimator.  It is used for the non-uniform layout comparison in Section~\ref{sec:nonuniform_layout_psf} and as a theoretical reference for the simulation results in Section~\ref{sec:simulation_A}.

Eliminating the unknown complex gain gives the delay CRB for an arbitrary
measured-frequency set as
\begin{align}
\mathrm{CRB}(\tau_0)
&=
\frac{\sigma^2}{8\pi^2|\alpha_0|^2}
\notag\\
&\quad\times
\left[
\sum_{t=1}^{T}\sum_{m\in\mathcal{U}} f_{t,m}^2
-
\frac{\left(\sum_{t=1}^{T}\sum_{m\in\mathcal{U}}f_{t,m}\right)^2}{TM_b}
\right]^{-1}.
\label{eq:crb_var_form_multiband}
\end{align}
The derivation is provided in Appendix~\ref{app:crb_single}. The CRB expression shows that the delay bound is inversely proportional to the second central moment of the measured frequencies.

We next express this frequency-variance term for a \(T\)-subband layout with
arbitrary subband-center frequencies. Using \(\sum_{m\in\mathcal{U}} m=0\), the first-
and second-order frequency sums are
\begin{align}
\sum_{t=1}^{T}\sum_{m\in\mathcal{U}} f_{t,m}
&=
M_b\sum_{t=1}^{T} F_t,
\label{eq:sum_f_multiband}
\\
\sum_{t=1}^{T}\sum_{m\in\mathcal{U}} f_{t,m}^2
&=
M_b\sum_{t=1}^{T} F_t^2
+
T\frac{M_b(M_b^2-1)}{12}(\Delta f)^2 .
\label{eq:sum_f2_multiband}
\end{align}
Substituting \eqref{eq:sum_f_multiband} and \eqref{eq:sum_f2_multiband} into
\eqref{eq:crb_var_form_multiband} gives
\begin{align}
&\sum_{t=1}^{T}\sum_{m\in\mathcal{U}} f_{t,m}^2
-
\frac{\left(\sum_{t=1}^{T}\sum_{m\in\mathcal{U}}f_{t,m}\right)^2}{TM_b}
\notag\\
&=
M_b\left(
\sum_{t=1}^{T} F_t^2
-
\frac{(\sum_{t=1}^{T} F_t)^2}{T}
\right)
+
T\frac{M_b(M_b^2-1)}{12}(\Delta f)^2 .
\label{eq:var_term_multiband_general_centers}
\end{align}
The first term is the inter-band contribution determined by the second central
moment of the subband centers, and the second term is the intra-band
contribution due to the finite bandwidth within each subband.

For the canonical equal-spaced subband-center layout in
\eqref{eq:centers_equal_spacing}, the inter-band contribution reduces to 
\begin{align}
\sum_{t=1}^{T} F_t^2
-
\frac{(\sum_{t=1}^{T} F_t)^2}{T}
&=
\frac{T(T^2-1)}{12}f_{\mathrm{gap}}^2 .
\label{eq:center_moment_closed}
\end{align}
Using \eqref{eq:var_term_multiband_general_centers} and
\eqref{eq:center_moment_closed} in \eqref{eq:crb_var_form_multiband} gives the single-path CRB for the canonical \(T\)-subband configuration as
\begin{align}
\mathrm{CRB}_{\mathrm{MB}}(\tau_0)
&=
\frac{\sigma^2}{8\pi^2|\alpha_0|^2\mathcal{V}_{\mathrm{MB}}},
\label{eq:crb_multiband_final}
\\
\mathcal{V}_{\mathrm{MB}}
&=
\frac{TM_b}{12}
\left(
(T^2-1)f_{\mathrm{gap}}^2
+
(M_b^2-1)(\Delta f)^2
\right) .
\notag
\end{align}
In \eqref{eq:crb_multiband_final}, \(\mathcal{V}_{\mathrm{MB}}\) is the
effective frequency-variance term in the CRB denominator. The factor \(TM_b\)
reflects the total number of CFR measurements. Inside the parentheses,
\((T^2-1)f_{\mathrm{gap}}^2\) is the inter-band term produced by the spread of
the equal-spaced subband centers, so increasing either \(T\) or
\(f_{\mathrm{gap}}\) strengthens this contribution. The term
\((M_b^2-1)(\Delta f)^2\) is the intra-band term produced by the finite
subband bandwidth, so increasing \(M_b\) or \(\Delta f\) strengthens this
contribution. A larger \(\mathcal{V}_{\mathrm{MB}}\) corresponds to a lower
single-path delay CRB.
}

\section{\texorpdfstring{Simulation results}{Simulation results}} \label{sec:simulation}

\polish{
To evaluate the proposed multi-band delay estimation framework, the numerical
configuration follows an IEEE~802.11be (Wi-Fi~7) setting in the \(6\)~GHz band.
The available spectrum supports the use of disjoint subbands over a large frequency span,
making it suitable for high-resolution delay estimation and localization
~\cite{WIFI7}. The measured frequency set \(\mathcal{F}\) contains uniformly
spaced frequencies within the selected disjoint subbands. The subband layout and channel
model are specified for each experiment.

Each Monte Carlo trial generates one multi-band CFR snapshot with additive
white Gaussian noise. All statistical results are computed over \(1000\)
independent trials. The PSO parameters are set to $N_p=15$, $Q=50$, $w_s=0.99$, $w_e=0.4$,
$c_1=1.5$, $c_2=0.5$, $\gamma=2$, and $\varepsilon=10^{-5}$. All execution-time results reported in Tables~\ref{tab:exec_time} and~\ref{tab:multipath_runtime}
were measured on the same workstation equipped with an AMD Ryzen~7 PRO~7840HS CPU and 32~GB RAM.}

\subsection{\texorpdfstring{\rev{Single-Path Delay Estimation}}{Single-Path Delay Estimation}}\label{sec:simulation_A}
\begin{figure}[t]
\centering
\includegraphics[width=0.7\linewidth]{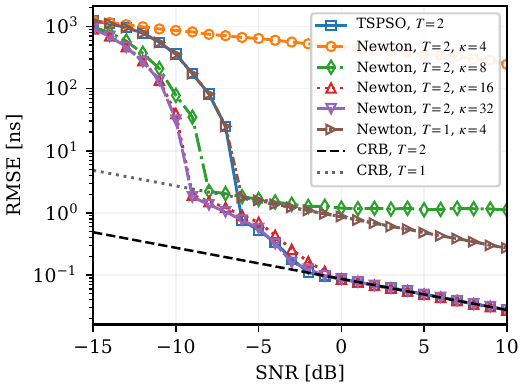}
\caption{\rev{Single-path delay RMSE versus SNR for TSPSO and Newton refinement with different coarse-grid oversampling factors $\kappa$. The multi-band layout uses $T=2$, $f_{\mathrm{gap}}=160$~MHz, and $M_b\Delta f=40$~MHz.}}
\label{fig:pso_rmse}
\end{figure}

\rev{
We first evaluate single-path delay estimation under the canonical equal-spaced subband-center layout. In each Monte Carlo trial, the delay $\tau_0$ is drawn uniformly from $100$ to $600$~ns. The path-gain magnitude is normalized, and the additive white Gaussian noise level is adjusted for each SNR point. The corresponding CRB is computed from the closed-form expression in~\eqref{eq:crb_multiband_final}.

Fig.~\ref{fig:pso_rmse} compares TSPSO with Newton refinement initialized by coarse
delay grids with different oversampling factors $\kappa$, where the
grid spacing is $\Delta_g/\kappa$.
The single-band Newton baseline uses only one $40$~MHz subband and therefore has a higher CRB. With $T=2$, the two $40$~MHz subbands are separated by $f_{\mathrm{gap}}=160$~MHz, which increases the effective bandwidth and lowers the CRB. 

For Newton refinement with $T=2$, the coarse-grid search is evaluated directly on the full multi-band objective. A small $\kappa$ may fail to sample the attraction basin of the mainlobe. Newton refinement can then converge to a sidelobe local maximum, producing an RMSE floor even at high SNR. Increasing $\kappa$ reduces this coarse-grid initialization error, but the grid-search cost increases rapidly as shown in Table~\ref{tab:exec_time}.

In contrast, TSPSO initializes the refinement interval using the
$40$~MHz reference-subband objective. Its smoother response avoids the
dense local maxima of the multi-band objective and reduces the
initialization cost. However, the single-band objective uses fewer
measurements and is therefore more sensitive to noise than the multi-band objective. At low SNR, noise can strongly perturb the reference-subband estimate and reduce the probability that the true
delay lies within the refinement interval. Consequently, TSPSO performs close to the single-band Newton baseline
with $\kappa=4$ in the low-SNR region. When the SNR
exceeds about $-6$~dB, the reference-subband estimate becomes
sufficiently reliable, and TSPSO approaches the RMSE of $T=2$ Newton
refinement with $\kappa=32$ while requiring substantially fewer grid
evaluations.
}

\begin{table}[t]
\centering
\caption{\rev{Execution time for Newton refinement with coarse-grid initialization under different oversampling factors $\kappa$, compared with TSPSO. The corresponding delay-grid spacing is $\Delta_{\mathrm{g}}/\kappa$, where $\Delta_{\mathrm{g}}=25$ ns.}}

\label{tab:exec_time}
\footnotesize
\setlength{\tabcolsep}{5pt}
\renewcommand{\arraystretch}{1.08}
\begin{tabular}{l c}
\toprule
\rev{Method / $\kappa$} & \rev{Execution time (ms)} \\
\midrule
\rev{Newton refinement, $\kappa=4$ ($T=2$)}        & \rev{4.55}   \\
\rev{Newton refinement, $\kappa=8$ ($T=2$)}       & \rev{8.35} \\
\rev{Newton refinement, $\kappa=16$ ($T=2$)}       & \rev{16.69} \\
\rev{Newton refinement, $\kappa=32$ ($T=2$)}       & \rev{33.61} \\
\rev{Newton refinement, $\kappa=4$ ($T=1$)}        & \rev{2.51} \\
\rev{TSPSO ($T=2$)}                                & \rev{7.13}  \\
\bottomrule
\end{tabular}
\end{table}

\begin{figure*}[t]
    \centering
    \begin{minipage}[t]{0.33\textwidth}
        \centering
        \includegraphics[width=\linewidth]{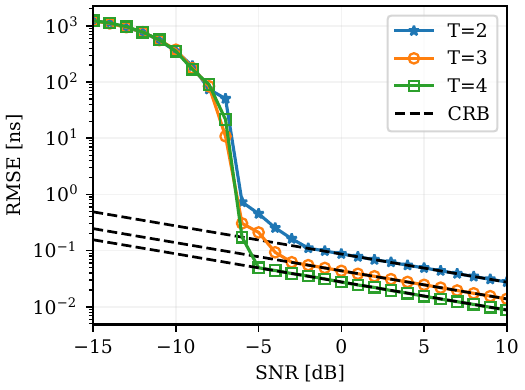}
        \vspace{-2mm}

        {\footnotesize\rev{(a)}}
    \end{minipage}\hfill
    \begin{minipage}[t]{0.33\textwidth}
        \centering
        \includegraphics[width=\linewidth]{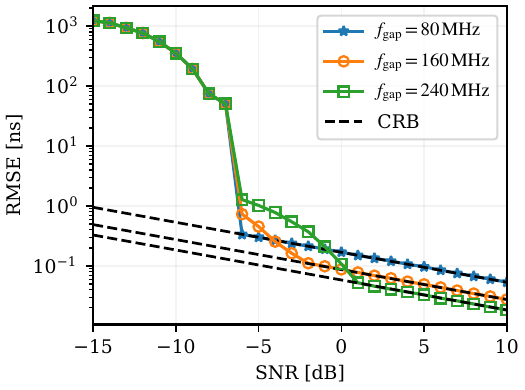}
        \vspace{-2mm}

        {\footnotesize\rev{(b)}}
    \end{minipage}\hfill
    \begin{minipage}[t]{0.33\textwidth}
        \centering
        \includegraphics[width=\linewidth]{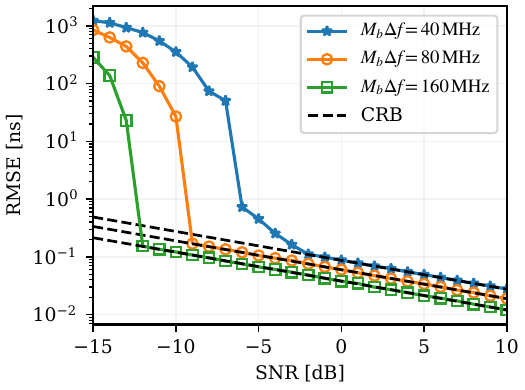}
        \vspace{-2mm}

        {\footnotesize\rev{(c)}}
    \end{minipage}

    \vspace{0mm}
\caption{\rev{TSPSO single-path delay RMSE versus SNR under canonical equal-spaced subband-center layouts, together with the corresponding CRB.
(a) Effect of the number of active subbands $T$ with $f_{\mathrm{gap}}=160$~MHz and $M_b\Delta f=40$~MHz.
(b) Effect of the frequency gap $f_{\mathrm{gap}}$ with $T=2$ and $M_b\Delta f=40$~MHz.
(c) Effect of the per-subband bandwidth $M_b\Delta f$ with $T=2$ and $f_{\mathrm{gap}}=160$~MHz.}}
    \label{fig:single_path_param_sweep}
    \vspace{-2mm}
\end{figure*}

\rev{
Fig.~\ref{fig:single_path_param_sweep} evaluates the TSPSO single-path delay RMSE under different canonical layout parameters. The CRB trends follow directly from~\eqref{eq:crb_multiband_final}. In Fig.~\ref{fig:single_path_param_sweep}a, increasing $T$ increases the inter-band frequency-variance term and therefore lowers the CRB. The RMSE decreases accordingly once the SNR is sufficiently high. Fig.~\ref{fig:single_path_param_sweep}b isolates the role of subband separation with fixed $T$.
A larger $f_{\mathrm{gap}}$ widens spread of the subband centers, so the CRB predicted by~\eqref{eq:crb_multiband_final} becomes lower. However, the canonical PSF in~\eqref{eq:psf_multiband_final} shows that the spacing between periodic sidelobe peaks is $1/f_{\mathrm{gap}}$. A larger frequency gap moves these peaks closer to $\delta_\tau=0$, where they are less attenuated by the intra-band Dirichlet envelope. At intermediate SNR, these strong sidelobe peaks can become competitive with the mainlobe in the noisy objective, increasing the probability that the estimator locks onto a sidelobe peak. As a result, larger-gap layouts require a higher SNR before the mainlobe dominates reliably and the RMSE approaches the CRB.

Finally, we fix \(\Delta f = 312.5\)~kHz and vary the \(M_b\). In this setting, increasing
\(M_b\) directly increases the per-subband bandwidth \(M_b\Delta f\). As shown
in Fig.~\ref{fig:single_path_param_sweep}c, a larger \(M_b\Delta f\) reduces
the RMSE and allows the estimator to approach the CRB at a lower SNR. A wider subband benefits the estimator in several ways. It increases the
intra-band frequency-variance term in \eqref{eq:crb_multiband_final}, thereby
lowering the CRB. It also gives the reference-subband coarse stage a narrower
objective mainlobe and better delay discrimination, making the coarse estimate
more robust at low SNR. In addition, the larger subband bandwidth narrows the
Dirichlet envelope in \eqref{eq:psf_multiband_final}, which suppresses
periodic sidelobe peaks more strongly and reduces delay ambiguity. Thus, increasing the per-subband bandwidth improves both local delay accuracy
and sidelobe robustness. This gain, however, comes with higher computational
cost because more frequency samples are processed in each subband.
}

\subsection{\texorpdfstring{\rev{Impact of Frequency Layout on LoS Estimation in Multipath Channels}}{Impact of Frequency Layout on LoS Estimation in Multipath Channels}}\label{sec:multipath_layout}

\begin{figure*}[t]
    \centering
    \begin{minipage}[t]{0.33\textwidth}
        \centering
        \includegraphics[width=\linewidth]{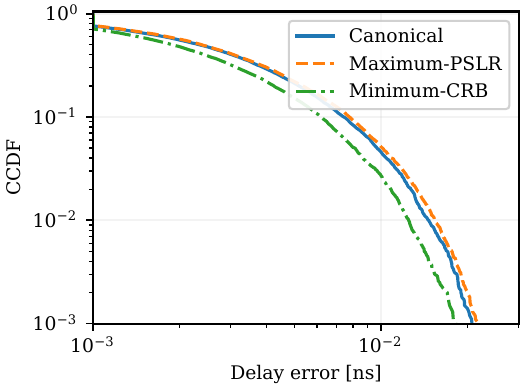}
        \vspace{-2mm}

        {\footnotesize\rev{(a)}}
    \end{minipage}\hfill
    \begin{minipage}[t]{0.33\textwidth}
        \centering
        \includegraphics[width=\linewidth]{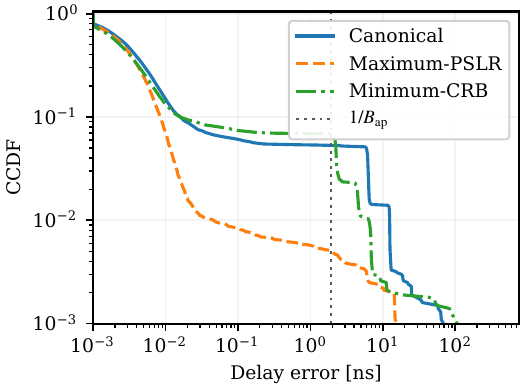}
        \vspace{-2mm}

        {\footnotesize\rev{(b)}}
    \end{minipage}\hfill
    \begin{minipage}[t]{0.33\textwidth}
        \centering
        \includegraphics[width=\linewidth]{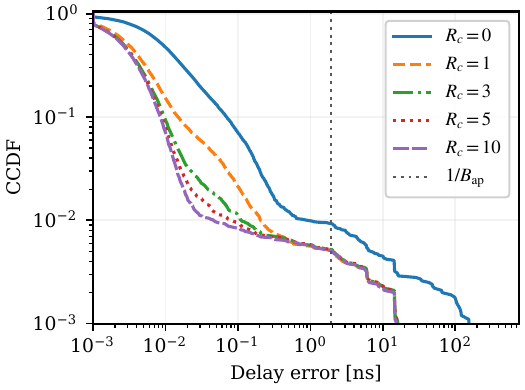}
        \vspace{-2mm}

        {\footnotesize\rev{(c)}}
    \end{minipage}

    \vspace{0mm}
\caption{\rev{LoS delay error CCDF for PSO-RELAX under the two five-path separation regimes.
(a) Well-separated regime (\(R_c=10\)).
(b) Closely spaced regime (\(R_c=10\)).
(c) Effect of RELAX cycles in the closely spaced regime under the Maximum-PSLR layout.}}
    \label{fig:multipath_layout_ccdf}
    \vspace{-2mm}
\end{figure*}

\rev{
We next evaluate LoS delay estimation in multipath channels under the three
representative equal-resource layouts in Fig.~\ref{fig:irregular_layout_psf}.
Unlike the single-path simulations, the multipath performance depends not only
on delay sensitivity but also on inter-path interference through the PSF
sidelobes. We begin with five-path channels. In each Monte Carlo realization, the
path delays are drawn uniformly from $[100,600]$~ns subject to a
minimum-spacing constraint. Let $B_{\mathrm{ref}}=M_b\Delta f=40$~MHz denote the
reference-subband bandwidth. The minimum spacing is set to
$2.5/B_{\mathrm{ref}}$ for the well-separated regime and
$0.1/B_{\mathrm{ref}}$ for the closely spaced regime. Across paths, the phases and SNRs are
independently drawn from uniform distributions over $[0,2\pi)$ and
$[10,25)$~dB, respectively. 

Fig.~\ref{fig:multipath_layout_ccdf} evaluates the LoS delay error complementary cumulative distribution function (CCDF) under the two separation regimes. In Fig.~\ref{fig:multipath_layout_ccdf}a, the paths are well separated and delay sensitivity dominates, so the Minimum-CRB layout gives the best LoS delay error distribution. In Fig.~\ref{fig:multipath_layout_ccdf}b, the paths are closely spaced, inter-path interference becomes more important, and layouts with larger PSF sidelobes are more strongly affected. We use $1/B_{\mathrm{ap}}$ as the aperture-resolution reference, where $B_{\mathrm{ap}}\triangleq \max_{t,m} f_{t,m}-\min_{t,m} f_{t,m}$ is the common end-to-end active frequency span. Since all three layouts have the same $B_{\mathrm{ap}}$, they share the same $1/B_{\mathrm{ap}}\approx1.92$~ns. A trial with LoS delay error below $1/B_{\mathrm{ap}}$ is counted as a successful aperture-resolution estimate. In the closely spaced regime, the Maximum-PSLR layout achieves more than $99\%$ success probability, while the canonical and Minimum-CRB layouts still exceed $90\%$ but are more affected by their larger sidelobes. Fig.~\ref{fig:multipath_layout_ccdf}c sweeps the number of RELAX cycles $R_c$ for the closely spaced regime under the Maximum-PSLR layout. The case $R_c=0$ corresponds to PSO-OMP, where no cyclic RELAX refinement is performed. Larger $R_c$ can reduce the LoS delay error, but the gain becomes marginal once $R_c>5$. In the error region above about $1$~ns, additional RELAX cycles provide little improvement because strong inter-path interference through PSF sidelobes becomes the limiting factor. Based on this saturation behavior, we set $R_c=10$ for PSO-RELAX in the subsequent multipath simulations.
}

\begin{figure*}[t]
    \centering
    \begin{minipage}[t]{0.33\textwidth}
        \centering
        \includegraphics[width=\linewidth]{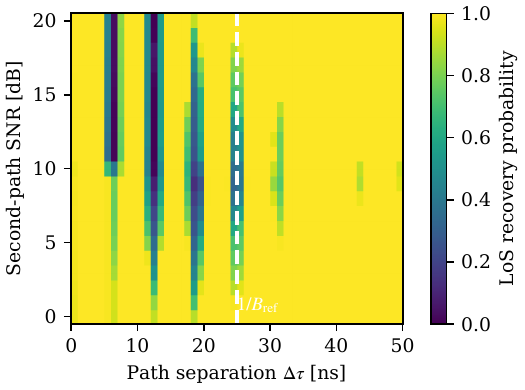}
        \vspace{-2mm}

        {\footnotesize\rev{(a)}}
    \end{minipage}\hfill
    \begin{minipage}[t]{0.33\textwidth}
        \centering
        \includegraphics[width=\linewidth]{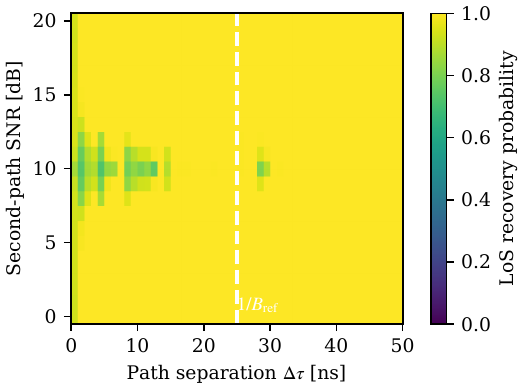}
        \vspace{-2mm}

        {\footnotesize\rev{(b)}}
    \end{minipage}\hfill
    \begin{minipage}[t]{0.33\textwidth}
        \centering
        \includegraphics[width=\linewidth]{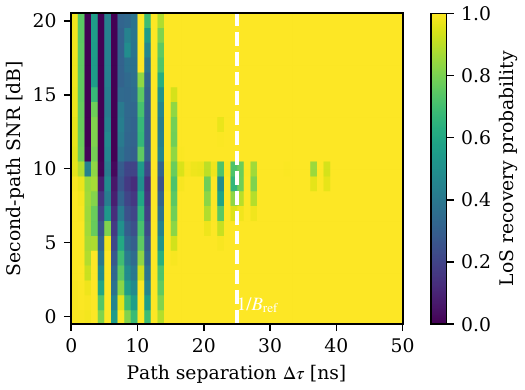}
        \vspace{-2mm}

        {\footnotesize\rev{(c)}}
    \end{minipage}

    \vspace{0mm}
\caption{\rev{LoS recovery probability in a controlled two-path near-far experiment.
(a) Canonical layout.
(b) Maximum-PSLR layout.
(c) Minimum-CRB layout.}}
    \label{fig:nearfar_layout_success}
    \vspace{-2mm}
\end{figure*}

\rev{
Fig.~\ref{fig:nearfar_layout_success} further studies how inter-path interference affects LoS estimation under different frequency layouts. We use a controlled two-path near-far experiment. The LoS path is fixed at $\tau_{\mathrm{LoS}}=66$~ns with SNR $10$~dB. A second path is placed at $\tau_{\mathrm{LoS}}+\Delta\tau$. We sweep the path separation $\Delta\tau$ and the second-path SNR. For each pair of path separation and second-path SNR, the LoS recovery probability is evaluated using the same $1/B_{\mathrm{ap}}$ success criterion.

The dependence on $\Delta\tau$ follows the PSF sidelobe structure. Low-recovery regions occur at delay separations that align with strong sidelobes of the corresponding PSF in Fig.~\ref{fig:irregular_layout_psf}c. This shows how a neighboring path can interfere with LoS estimation through the PSF sidelobes. For the canonical and Minimum-CRB layouts, the strong periodic sidelobes lead to periodic LoS recovery failures as $\Delta\tau$ varies, especially when the neighboring path is sufficiently strong. In contrast, the Maximum-PSLR layout suppresses the dominant sidelobes, so no clear periodic failure pattern is observed and the LoS recovery probability remains high over most of the near-far region. The remaining failures mainly occur when the two paths are closely spaced and have comparable SNRs, where the two components are more difficult to separate. When the path powers are clearly different, the LoS path can still be recovered reliably even if it is weaker than the neighboring path.
}
\subsection{\texorpdfstring{\rev{Robustness to Inter-Band Non-Coherence}}{Robustness to Inter-Band Non-Coherence}}
\label{sec:phase_offset_sim}

\begin{figure*}[t]
    \centering
    \begin{minipage}[t]{0.33\textwidth}
        \centering
        \includegraphics[width=\linewidth]{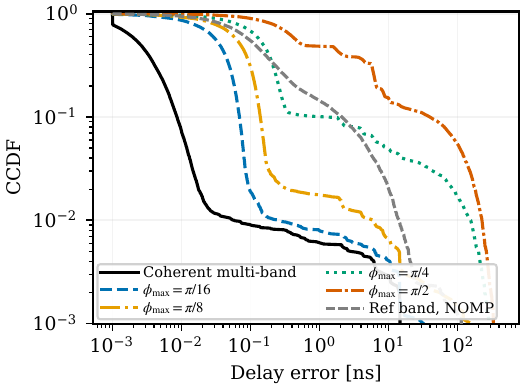}
        \vspace{-2mm}

        {\footnotesize\rev{(a)}}
    \end{minipage}\hfill
    \begin{minipage}[t]{0.33\textwidth}
        \centering
        \includegraphics[width=\linewidth]{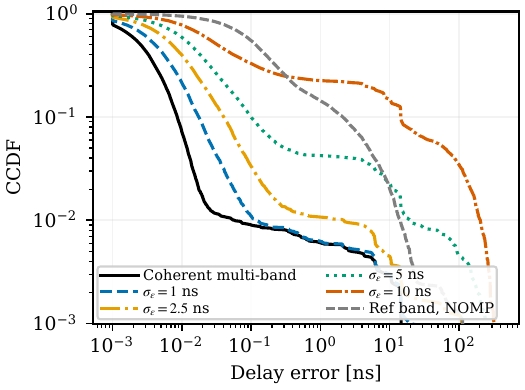}
        \vspace{-2mm}

        {\footnotesize\rev{(b)}}
    \end{minipage}\hfill
    \begin{minipage}[t]{0.33\textwidth}
        \centering
        \includegraphics[width=\linewidth]{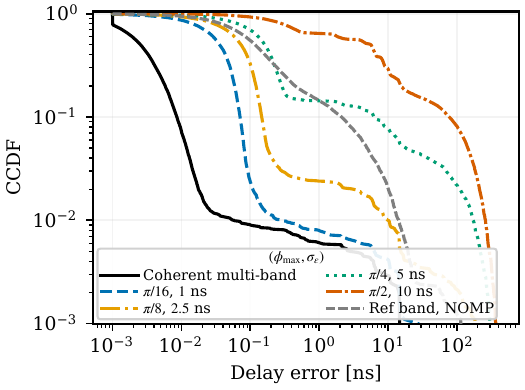}
        \vspace{-2mm}

        {\footnotesize\rev{(c)}}
    \end{minipage}

    \vspace{0mm}
\caption{\rev{LoS delay error CCDF under inter-band non-coherence. Results are obtained using PSO-RELAX with the Maximum-PSLR layout in the closely spaced five-path scenario.
(a) Effect of \(\phi_t\).
(b) Effect of \(\epsilon_t\).
(c) Joint effect of \(\phi_t\) and \(\epsilon_t\).}}
    \label{fig:phase_offset_calibration}
    \vspace{-2mm}
\end{figure*}

\rev{
The phase- and timing-offset effects characterized in
Section~\ref{sec:interband_noncoherence} are evaluated for PSO-RELAX using the
closely spaced five-path setting and the Maximum-PSLR layout. The coherent
multi-band baseline is
obtained by setting all relative offsets to zero.
NOMP with \(\kappa=4\) applied to the \(40\)~MHz reference subband provides a
single-subband baseline. 

Fig.~\ref{fig:phase_offset_calibration}a isolates the effect of
$\phi_t$ by setting $\epsilon_t=0$. For each non-reference subband,
$\phi_t$ is drawn uniformly from $[-\phi_{\max},\phi_{\max}]$.
The performance remains close to the coherent multi-band baseline for
$\phi_{\max}=\pi/16$. When $\phi_{\max}$ reaches or exceeds $\pi/4$,
the high-error tail extends beyond the single-subband baseline. As
$\phi_{\max}$ increases, the subband contributions associated with the
same physical path become less phase-aligned, as described by
\eqref{eq:noncoherent_delay_projection}. This reduces the dominance of
the peak associated with the true path and changes the relative heights
of nearby sidelobe and inter-path peaks. Since path detection selects
the maximum of \eqref{eq:omp_objective}, the reduced objective margin
makes the delay estimate more susceptible to sidelobe peaks and
inter-path interference.

Fig.~\ref{fig:phase_offset_calibration}b isolates the effect of
$\epsilon_t$ by setting $\phi_t=0$. For each non-reference subband, the
relative timing offset $\epsilon_t$ is independently drawn from a
Gaussian distribution,
$\epsilon_t\sim\mathcal{N}(0,\sigma_\epsilon^2)$~\cite{Splicer}, where
$\sigma_\epsilon$ denotes the standard deviation of the timing
synchronization error. The result for $\sigma_\epsilon=1$~ns remains
close to the coherent multi-band baseline. When $\sigma_\epsilon$
reaches or exceeds $5$~ns, the high-error tail extends beyond the
single-subband baseline. A larger $\sigma_\epsilon$ increases the spread
of the timing offsets across subbands. As a result, different subbands
favor different candidate delays for the same physical path. This
reduces the coherent gain at the true delay $\tau_\ell$ and can make
nearby sidelobe peaks more competitive.

Fig.~\ref{fig:phase_offset_calibration}c evaluates the case where both
relative phase and timing offsets are present. The error tails become
heavier as the offset levels increase, showing that residual inter-band
non-coherence degrades the coherent multi-band gain. These
results highlight the importance of inter-band offset calibration for
preserving the robustness of multi-band delay estimation. 
}

\subsection{\texorpdfstring{\rev{Performance Comparison}}{Performance Comparison}}
\label{sec:performance_comparison}

\begin{figure*}[t]
    \centering

    \begin{minipage}[t]{0.32\textwidth}
        \centering
        \includegraphics[width=\linewidth]{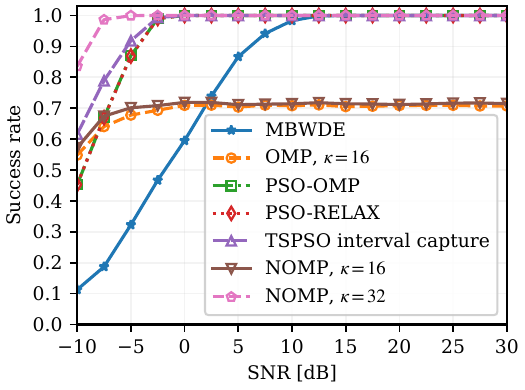}
        \vspace{-1mm}
        {\footnotesize\rev{(a)}}
        \vspace{-1mm}
    \end{minipage}\hfill
    \begin{minipage}[t]{0.32\textwidth}
        \centering
        \includegraphics[width=\linewidth]{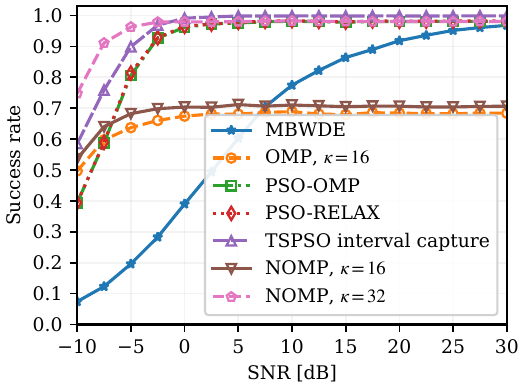}
        \vspace{-1mm}
        {\footnotesize\rev{(b)}}
        \vspace{-1mm}
    \end{minipage}\hfill
    \begin{minipage}[t]{0.32\textwidth}
        \centering
        \includegraphics[width=\linewidth]{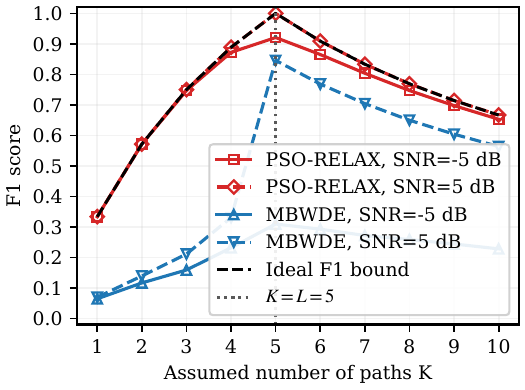}
        \vspace{-1mm}
        {\footnotesize\rev{(c)}}
        \vspace{-1mm}
    \end{minipage}

    \vspace{1.5mm}
\caption{\rev{Performance comparison under the Maximum-PSLR layout.
(a) Path-recovery success rate in the well-separated regime.
(b) Path-recovery success rate in the closely spaced regime.
(c) Path F1 score under assumed model-order mismatch in the well-separated
regime with true path count \(L=5\).
A path is counted as recovered when its delay error is below \(1/B_{\mathrm{ap}}\).}}
    \label{fig:COMPARE_all}
    \vspace{-2mm}
\end{figure*}

\rev{

In this subsection, Fig.~\ref{fig:COMPARE_all}a and
Fig.~\ref{fig:COMPARE_all}b compare the proposed PSO-OMP and
PSO-RELAX with discrete OMP~\cite{hisac}, the subspace-based MBWDE
method~\cite{subspace2}, and NOMP~\cite{NOMP} under the Maximum-PSLR
layout. OMP estimates the path delays on a discrete grid. NOMP uses the
same grid-based search to obtain initial delay estimates and then refines
them through Newton updates. Two oversampling factors are considered for
NOMP to illustrate the tradeoff between estimation accuracy and
computational cost.

The evaluation considers five-path channels under two path-separation
regimes. Fig.~\ref{fig:COMPARE_all}a uses the well-separated regime,
whereas Fig.~\ref{fig:COMPARE_all}b uses the closely spaced regime
defined in Section~\ref{sec:multipath_layout}. The LoS-path SNR is varied
from $-10$ to $30$~dB, and the SNRs of the remaining NLoS paths are set
to $(-3,-4.5,-6,-7)$~dB relative to the LoS path. The path-recovery success rate is defined as the fraction of true paths
recovered with a delay error below $1/B_{\mathrm{ap}}$. For
TSPSO, the interval-capture rate is also reported as the fraction of
true delays that fall within the refinement intervals constructed from
the first-stage reference-subband estimates. 

OMP exhibits a clear high-SNR ceiling because its estimates remain constrained to the \(\kappa=16\) delay grid. NOMP mitigates this basis mismatch through Newton refinement, but with \(\kappa=16\), the initial grid search may still select a sidelobe peak and the local Newton step may converge to this incorrect maximum. Increasing the oversampling factor to \(\kappa=32\) therefore gives the strongest NOMP baseline, at the cost of the highest computational
complexity among the compared methods. The low-SNR gap between NOMP with $\kappa=32$ and the PSO-based methods
extends the initialization tradeoff observed in the single-path results
of Fig.~\ref{fig:pso_rmse} to the multipath setting. NOMP performs a fine grid search on
the full $T=4$ multi-band residual, whereas TSPSO constructs its
refinement interval from the $T=1$ reference-subband residual. At low
SNR, the reference-subband residual is more strongly affected by noise,
which lowers the interval-capture rate. As the SNR increases, the
interval-capture rate rises rapidly in both path-separation regimes. In the
closely spaced regime, neighboring paths further perturb the
reference-subband residual, so reliable interval construction requires
a higher SNR. Above $5$~dB, the interval-capture rate approaches one in both path-separation regimes.

Once the refinement intervals contain the true delays, the PSO-based
estimators approach the performance of NOMP with $\kappa=32$. The
performance difference between the two path-separation regimes is then mainly
caused by sidelobe-induced inter-path interference. In the
well-separated regime, weak inter-path interference allows the recovery
rates to approach one. In the closely spaced regime, inter-path interference can bias the refined delay estimates and reduce the recovery rate, as illustrated in
Fig.~\ref{fig:nearfar_layout_success}. Nevertheless, both PSO-OMP and
PSO-RELAX achieve recovery rates above $95\%$ in the closely spaced
regime when the SNR exceeds $5$~dB. MBWDE also improves with SNR. However, in the single-snapshot setting,
noise can strongly perturb its estimated signal subspace. MBWDE
therefore requires a higher SNR to achieve accurate high-resolution
delay estimation and is less competitive than the sparse recovery methods in
the low-SNR region.

Table~\ref{tab:multipath_runtime} reports the average execution time for the
path-recovery experiments in Figs.~\ref{fig:COMPARE_all}a
and~\ref{fig:COMPARE_all}b. NOMP with \(\kappa=32\) provides the strongest
baseline recovery performance, but requires more than three times the runtime
of PSO-RELAX. By contrast, PSO-OMP and PSO-RELAX approach the recovery
performance of high-oversampling NOMP while remaining in the same runtime
range as OMP and MBWDE, demonstrating a favorable accuracy-complexity tradeoff.
}

\begin{table}[t]
\centering
\caption{\rev{Average execution time per five-path realization for the methods compared in Fig.~\ref{fig:COMPARE_all}a and Fig.~\ref{fig:COMPARE_all}b.}}
\label{tab:multipath_runtime}
\footnotesize
\setlength{\tabcolsep}{6pt}
\renewcommand{\arraystretch}{1.08}
\begin{tabular}{l c}
\toprule
\rev{Method} & \rev{Execution time (ms)} \\
\midrule
\rev{MBWDE} & \rev{73.187} \\
\rev{OMP, $\kappa=16$} & \rev{76.305} \\
\rev{PSO-OMP} & \rev{81.962} \\
\rev{PSO-RELAX} & \rev{92.575} \\
\rev{NOMP, $\kappa=16$} & \rev{160.263} \\
\rev{NOMP, $\kappa=32$} & \rev{305.046} \\
\bottomrule
\end{tabular}
\end{table}

\rev{
Fig.~\ref{fig:COMPARE_all}c evaluates the robustness of PSO-RELAX to
model-order mismatch in the well-separated five-path channel under the
Maximum-PSLR layout, where \(K\) denotes the assumed model order. The path F1 score is defined
as
}
\rev{\begin{align}
\mathrm{F1}
&=
\frac{2N_{\mathrm{detected}}}{L+K},
\label{eq:path_f1_score}
\end{align}}
\rev{where \(N_{\mathrm{detected}}\) is the number of detected paths whose
delay errors are below \(1/B_{\mathrm{ap}}\). Under perfect detection, the
corresponding ideal upper bound is \(2\min(K,L)/(K+L)\). When the model order is
underestimated, PSO-RELAX tends to extract the high-SNR paths first, so
its F1 score closely follows this ideal bound. When the model order is
overestimated, the additional estimates may be attracted by noise, sidelobes,
or residual leakage from already detected paths. This degradation is more
visible at low SNR. Nevertheless, PSO-RELAX remains substantially more robust than MBWDE across the tested values of \(K\).
}

\section{\texorpdfstring{\rev{Real-Data Measurements}}{Real-Data Measurements}}
\label{sec:real_data}

\begin{figure*}[t]
    \centering

    \begin{minipage}[t]{0.32\textwidth}
        \centering
        \includegraphics[width=\linewidth]{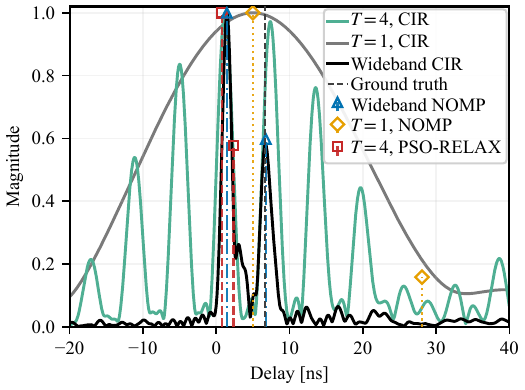}
        \vspace{-1mm}
        {\footnotesize\rev{(a) Canonical layout}}
        \vspace{-1mm}
    \end{minipage}\hfill
    \begin{minipage}[t]{0.32\textwidth}
        \centering
        \includegraphics[width=\linewidth]{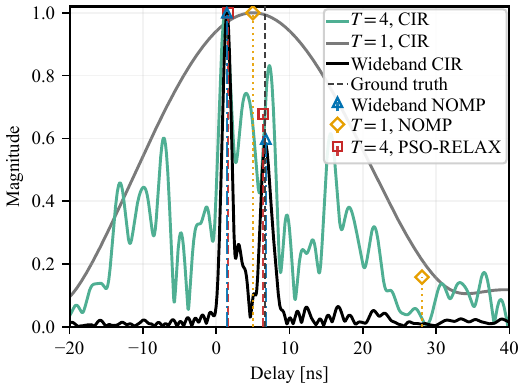}
        \vspace{-1mm}
        {\footnotesize\rev{(b) Maximum-PSLR layout}}
        \vspace{-1mm}
    \end{minipage}\hfill
    \begin{minipage}[t]{0.32\textwidth}
        \centering
        \includegraphics[width=\linewidth]{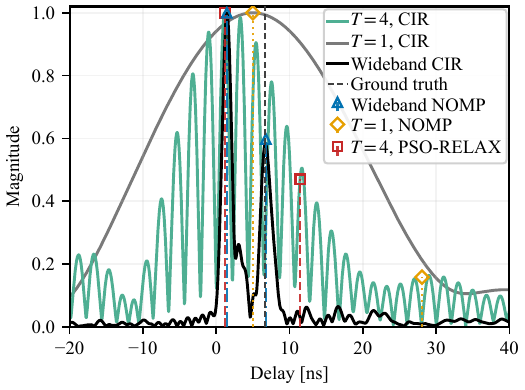}
        \vspace{-1mm}
        {\footnotesize\rev{(c) Minimum-CRB layout}}
        \vspace{-1mm}
    \end{minipage}

    \vspace{1.5mm}
    \caption{\rev{Representative measured CIRs and delay estimates for the
(a) canonical, (b) Maximum-PSLR, and (c) Minimum-CRB frequency layouts.
The \(T=1\) CIR and NOMP estimates use one \(40\)~MHz subband, whereas
the \(T=4\) CIR and PSO-RELAX results use four \(40\)~MHz
subbands. The Wideband CIR and Wideband NOMP estimates are obtained from
the full \(1\)~GHz CFR.}}
    \label{fig:measurement_representative}
    \vspace{-2mm}
\end{figure*}

\begin{figure}[H]
    \centering
    \includegraphics[width=0.65\linewidth]{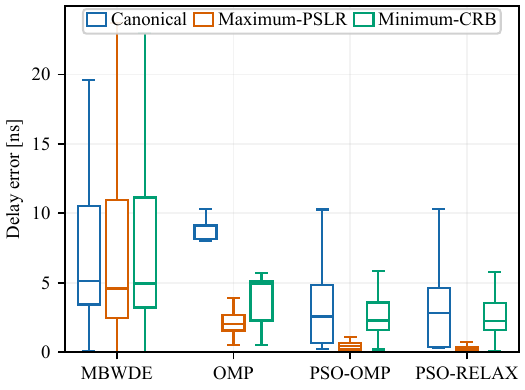}
    \caption{\rev{Box plots of the human-reflection delay errors for different estimators
under the canonical, Maximum-PSLR, and Minimum-CRB frequency layouts.}}
    \label{fig:measurement_statistics}
\end{figure}

\rev{
In this section, we evaluate the proposed estimators using experimental
CFR measurements from the dataset in~\cite{Pegoraro2025FR3}. The
measurements were collected in a scene containing a human target using a
coherent monostatic transceiver. Each acquisition provides a CFR snapshot over a \(1\)~GHz bandwidth centered at \(6.5\)~GHz, with a frequency spacing of \(1\)~MHz. From each snapshot, four \(40\)~MHz subbands are selected
according to the canonical, Maximum-PSLR, and Minimum-CRB layouts in
Fig.~\ref{fig:irregular_layout_psf}b. Since the subbands are selected
from the same wideband CFR snapshot, their relative phase is preserved. This
allows the frequency layouts to be evaluated under coherent multi-band
conditions.

NOMP is applied to one \(40\)~MHz subband as the
low-resolution \(T=1\) baseline and to the full \(1\)~GHz CFR to obtain
the high-resolution Wideband-NOMP reference estimates. The human-target distance measured with a
laser telemeter is converted to the corresponding round-trip propagation
delay and used as the ground truth. Because the laser reference point
and the dominant RF scattering point on the human body do not exactly
coincide, the measured delays exhibit a systematic offset. We estimate
this offset as the median error of the Wideband NOMP estimates compared to the ground truth and eliminate it from the all estimates.

Fig.~\ref{fig:measurement_representative} shows a representative
snapshot. The Wideband CIR resolves two dominant paths corresponding to
the direct Tx--Rx path and the human reflection. The
\(T=1\) CIR is limited by \(40\)~MHz bandwidth.
As a result, the two paths are merged into a single broad peak, and the \(T=1\) NOMP
estimates are biased. Using four non-contiguous subbands enlarges the effective bandwidth and improves delay resolution, but also produces layout-dependent sidelobes. In Fig.~\ref{fig:measurement_representative}a and Fig.~\ref{fig:measurement_representative}c, the path separation approximately coincides with a strong sidelobe offset of the corresponding PSF. The resulting inter-path leakage substantially biases the estimated human-reflection delay. This behavior agrees with the controlled two-path results in Fig.~\ref{fig:nearfar_layout_success}. In comparison, the Maximum-PSLR layout suppresses the dominant sidelobes and produces estimates closer to the Wideband reference.

Fig.~\ref{fig:measurement_statistics} reports the absolute human-reflection delay errors for 1600 CFR snapshots from 50 measurement records, each containing 32 repeated acquisitions. PSO-OMP and PSO-RELAX achieve lower median errors and tighter error distributions than MBWDE across all three layouts. The broader MBWDE distributions are consistent with the noise sensitivity of single-snapshot subspace estimation and indicate the advantage of sparse recovery when only limited measurements are available.

All three sparse recovery methods remain sensitive to the frequency layout and generally perform best with the Maximum-PSLR layout. OMP has larger errors because it is affected by both sidelobe ambiguity and basis mismatch from its discrete delay grid. The PSO-based continuous refinement in PSO-OMP and PSO-RELAX mitigates basis mismatch and improves robustness to sidelobe ambiguity. Among all evaluated configurations, PSO-RELAX with the Maximum-PSLR layout achieves the lowest median error.

These measurements show that PSF sidelobes directly affect the robustness of sparse delay estimation. Frequency-layout design should therefore consider both the CRB-predicted delay sensitivity and the sidelobe ambiguity of the PSF.
}

\section{Conclusion}

\polish{This paper investigated coherent multi-band ToA estimation from a
single CFR snapshot under non-contiguous frequency layouts. A multi-band
LS formulation, together with PSF and CRB analyses, was developed to
characterize delay sensitivity and sidelobe ambiguity. Based on these
insights, a PSO-assisted framework was proposed for  single-path
and sparse multipath delay estimation. In the tested multipath
scenarios, PSO-RELAX outperformed discrete OMP and MBWDE and approached
the performance of NOMP at substantially lower computational cost.
Real measurements further showed that layouts with lower PSF sidelobes
provided more robust delay estimates, highlighting the importance of
frequency-layout design. Overall, the proposed framework provides a
practical solution for high-resolution multi-band ToA estimation in
future wireless localization systems.}


\appendices

\section{Derivation of the Single-Path Delay CRB}
\label{app:crb_single}
\polish{
Consider the single-path observation model in
\eqref{eq:vec_model_single}, with mean
\(\boldsymbol{\mu}=\alpha_0\mathbf{a}(\tau_0)\) and real-valued parameter
vector
\(\boldsymbol{\theta}=
[\Re\{\alpha_0\},\Im\{\alpha_0\},\tau_0]^\top\).
The \((p,q)\)th entry of the Fisher information matrix is
\begin{equation}
[\mathbf{J}(\boldsymbol{\theta})]_{p,q}
=
\frac{2}{\sigma^2}
\Re\!\left\{
\left(\frac{\partial\boldsymbol{\mu}}{\partial\theta_p}\right)^{\!H}
\left(\frac{\partial\boldsymbol{\mu}}{\partial\theta_q}\right)
\right\}.
\label{eq:single_path_fim_app}
\end{equation}
Let \(\mathbf{D}\) denote the diagonal matrix containing the measured
frequencies \(f_{t,m}\). The required derivatives are
\(\partial\boldsymbol{\mu}/\partial\Re\{\alpha_0\}=\mathbf{a}(\tau_0)\),
\(\partial\boldsymbol{\mu}/\partial\Im\{\alpha_0\}=j\mathbf{a}(\tau_0)\),
and
\(\partial\boldsymbol{\mu}/\partial\tau_0
=-j2\pi\alpha_0\mathbf{D}\mathbf{a}(\tau_0)\).
Eliminating the complex gain by the Schur complement gives
\begin{align}
J_{\tau_0}
&=
\frac{8\pi^2|\alpha_0|^2}{\sigma^2}
\left[
\mathbf{a}(\tau_0)^H\mathbf{D}^2\mathbf{a}(\tau_0)
-
\frac{\left|\mathbf{a}(\tau_0)^H\mathbf{D}\mathbf{a}(\tau_0)\right|^2}
{\mathbf{a}(\tau_0)^H\mathbf{a}(\tau_0)}
\right].
\notag\\
&=
\frac{8\pi^2|\alpha_0|^2}{\sigma^2}
\left[
\sum_{t=1}^{T}\sum_{m\in\mathcal{U}}f_{t,m}^2
-
\frac{\left(\sum_{t=1}^{T}\sum_{m\in\mathcal{U}}f_{t,m}\right)^2}{TM_b}
\right],
\label{eq:single_path_delay_information_app}
\end{align}
where the second equality uses
\(\mathbf{a}(\tau_0)^H\mathbf{a}(\tau_0)=TM_b\).
Thus, \(\mathrm{CRB}(\tau_0)=J_{\tau_0}^{-1}\), which gives
\eqref{eq:crb_var_form_multiband}.
}

\section{Properties of the Dirichlet Kernel}
\label{app:dirichlet_props}

Let $\mathcal{U}=\left\{-\tfrac{M_b-1}{2},\ldots,\tfrac{M_b-1}{2}\right\}$ with odd $M_b$ and define
\begin{equation}
D(\Delta)
\triangleq
\sum_{m\in\mathcal{U}} e^{-j2\pi m\Delta f\,\Delta}
=
\frac{\sin(\pi M_b\Delta f\,\Delta)}{\sin(\pi \Delta f\,\Delta)}.
\label{eq:D_app}
\end{equation}
Since $D(\Delta)$ is periodic in $\Delta$ with period $1/\Delta f$, we restrict to the alias-free interval
$|\Delta|<1/(2\Delta f)$.

\medskip
\noindent\textbf{Lemma~1.}
Within $|\Delta|<1/(2\Delta f)$, $|D(\Delta)|$ attains its unique global maximum at $\Delta=0$.
Moreover, $D(\Delta)$ is strictly decreasing for $\Delta\in(0,\,1/(2M_b\Delta f)]$ and, for any
$\Delta\in(1/(2M_b\Delta f),\,1/(2\Delta f))$, \(|D(\Delta)|
<
D\!\left(\frac{1}{2M_b\Delta f}\right).\)

\medskip
\noindent\textit{Proof.}
Define $u\triangleq \Delta f\,\Delta$. 
For $u\in(0,1/(2M_b)]$, differentiating \eqref{eq:D_app} gives
$\frac{\mathrm{d}D}{\mathrm{d}u}\propto M_b\tan(\pi u)-\tan(M_b\pi u)$, which is strictly negative because
$x\mapsto\tan(x)/x$ is strictly increasing on $(0,\pi/2)$. For $u\in(1/(2M_b),1/2)$ we have $\pi u\in(\pi/(2M_b),\pi/2)$, and $\sin(x)$ is strictly increasing on
$(0,\pi/2)$. Hence $\sin(\pi u)>\sin(\pi/(2M_b))$. Moreover, $|\sin(M_b\pi u)|\le 1$, which yields \(|D(\Delta)|
=\left|\frac{\sin(M_b\pi u)}{\sin(\pi u)}\right|
\le \frac{1}{\sin(\pi u)}
< \frac{1}{\sin(\pi/(2M_b))}
= D\!\left(\frac{1}{2M_b\Delta f}\right).\)

\hfill\qedhere

\bibliographystyle{IEEEtran}
\bibliography{multiband}


 




\end{document}